\documentclass[epj,lineno,final]{svjour}    

\usepackage[hyperindex,breaklinks]{hyperref} 
  
\usepackage[switch]{lineno}    
\def\linenumberfont{\footnotesize}         
    
\usepackage{atlasphysics}     

\usepackage{epsfig}  
\usepackage{graphics}
\usepackage{graphicx}
  
\usepackage{wasysym}
\usepackage{lscape} 
\usepackage{color}  
          
\usepackage[noadjust]{cite}

\usepackage[section]{placeins}

\def\beq{\begin{equation}}
\def\eeq{\end{equation}}
\def\bec{\begin{center}}
\def\eec{\end{center}}
\def\bef{\begin{figure*}}
\def\eef{\end{figure*}}
\def\ct{\cite}

\def\pT{p_{\rm T}}
\def\kT{k_{\rm T}}

\usepackage{preprintcover}

\PreprintCoverPaperTitle{Two-particle Bose--Einstein correlations in $pp$ collisions at $\mathbf {\sqrt{s} =}$ 0.9 and 7~TeV
measured with the ATLAS detector 
}  

\PreprintIdNumber{CERN-PH-EP-2014-264}  

\PreprintCoverAbstract{
The paper   presents studies of Bose--Einstein Correlations (BEC) for pairs of   like-sign  charged particles measured in the kinematic  range $\pT >$~100~\MeV\ and $|\eta|<$~2.5  in   proton--proton  collisions at centre-of-mass energies of 0.9  and 7~\TeV\ with the ATLAS detector at the CERN Large Hadron Collider. The integrated luminosities are  approximately
 7~$\mu$b$^{-1}$,  190~$\mu$b$^{-1}$  and   12.4~nb$^{-1}$  for   0.9~\TeV,  7~\TeV\ minimum-bias  and 7~\TeV\  high-multiplicity  data samples,   respectively. The multiplicity dependence of the BEC  parameters characterizing the  correlation strength  and the correlation source size  are investigated for charged-particle  multiplicities of up to 240.  A saturation effect in the multiplicity dependence of the correlation   source size   parameter   is observed  using the  high-multiplicity   7~\TeV\ data sample.  The dependence of the BEC parameters  on the average transverse momentum of the  particle pair  is  also   investigated.   
}

\PreprintJournalName{Eur.\ Phys.\ J.\ C {\bf 75} (2015) 466}

\begin{document}

\twocolumn
\title{Two-particle Bose--Einstein correlations in $pp$ collisions at $\mathbf {\sqrt{s} =}$  0.9 and 7 TeV
measured with the ATLAS detector}
\author{\bf The ATLAS Collaboration 
}
\institute{
CERN, 1211 Geneva 23, Switzerland
}

\abstract{
The paper   presents studies of Bose--Einstein Correlations (BEC) for pairs of   like-sign  charged particles
 measured in the kinematic  range $\pT >$~100~\MeV\ and $|\eta|<$~2.5  in   proton--proton  
 collisions at centre-of-mass energies of 0.9  and 7~\TeV\ with the ATLAS detector at the CERN Large Hadron Collider. 
The integrated luminosities are  approximately 7~$\mu$b$^{-1}$,  
190~$\mu$b$^{-1}$  and   12.4~nb$^{-1}$ 
 for   0.9~\TeV,  7~\TeV\ minimum-bias  and 7~\TeV\  high-multiplicity  data samples,   respectively.
The multiplicity dependence of the BEC  parameters characterizing the  correlation strength 
and the correlation source size  are investigated for charged-particle 
multiplicities of up to 240. 
A saturation effect in the multiplicity dependence of the correlation 
source size 
parameter 
is observed  using the  high-multiplicity  
7~\TeV\ data sample.  The dependence of the BEC parameters  on the average transverse momentum of the 
particle pair  is  also   investigated.   
\\[2mm]
} 

\date{   \mbox{~~} }

\authorrunning{The ATLAS Collaboration}
\titlerunning{Two-particle Bose--Einstein correlations}
\maketitle

\section{Introduction}
\label{intro}

Particle correlations play an important role in the understanding of multiparticle production.
Correlations between identical bosons, called Bose-Einstein correlations (BEC),  are a well-known  
 phenomenon in high-energy and nuclear physics
(for 
reviews see 
\ct{becrev-zajc,hadrorev-jap,hadrorev-jap-2,Boal:1990yh,Lisa:2005dd,becrev-kit,becrev-tamas,becrev-alex,book,becrev-weiner,becrev-weiner2,qo-jap}).
The BEC are often  considered to be  the  analogue of the Hanbury-Brown and Twiss effect  \ct{hbt,hbt2,hbt3}  in astronomy, 
describing the interference of incoherently emitted identical bosons \ct{hbt2bec,hbt2bec2,hbt2bec-1,hbt2bec-2}. 
 They represent    a sensitive   probe of the space-time geometry of the hadronization 
 region and allow   the determination of  the size and the shape of the source from which  particles   are emitted.

The production of identical bosons 
that  are  close together in phase space is enhanced by the presence of BEC.
The first observation of BEC effects
in identically charged   pions produced in $p\bar{p}$ collisions was 
reported  in  Ref.\ \ct{gglp,gglp2}. 
Since then, BEC have been studied for systems of two or more identical bosons produced in 
 various types of collisions, from
 leptonic to hadronic and nuclear collisions 
(see  Refs.\ \ct{becrev-zajc,hadrorev-jap,hadrorev-jap-2,Boal:1990yh,Lisa:2005dd,becrev-tamas,becrev-kit,becrev-alex,book} and references therein). 

Studies of the dependence of BEC
on particle multiplicity and transverse momentum are of special interest.
They help  to understand the multiparticle production mechanism. 
 The size of the source emitting the correlated particles has  been observed to increase with particle multiplicity. 
This can be understood as  arising from the  increase in the initial 
geometrical  region of overlap of the colliding objects \ct{overlap}:
a large overlap implies a large multiplicity. 
 While this dependence is natural in nucleus--nucleus collisions, 
 the increase of size with multiplicity has also been observed in hadronic and leptonic 
 interactions.  In  the latter, it is understood as a result of 
superposition of many sources \ct{becrev-alex,bec-mult,bec-mult-1,bec-mult-12,bec-mult-2,bec-mult-22}
 or  related to the number of jets \ct{opal-BEC-mult,alex-mult}.
High-multiplicity data in proton-proton  interactions can  serve as a  reference   
for studies of nucleus--nucleus  collisions. 
 The effect is reproduced in both the hydrodynamical/hydrokinetic   
\ct{sinyukov,sinyukov-1,sinyukov-2} and 
 Pomeron-based \ct{schegelsky,schegelsky-1} approaches  for hadronic  interactions where high
multiplicities play a  crucial role. 
The dependence on the transverse momentum of the emitter particle pair
is another important feature of the BEC effect \ct{shuryak-hic}. 
 In nucleus--nucleus collisions the dependence  of the particle emitter 
size on the transverse momentum is explained as a ``collective flow'', 
which generates a characteristic fall-off of the emitter size with 
increasing  transverse momentum \ct{kT-th,kT-th-1,Hama:1987xv}
while strong space-time   momentum-energy  correlations 
 may offer an explanation in  more ``elementary'' leptonic and hadronic 
systems 
 \ct{shuryak-hic,becrev-kit,becrev-tamas,book,sinyukov,sinyukov-1,sinyukov-2} 
where   BEC measurements   serve as a  test of different models 
\ct{sinyukov,sinyukov-1,sinyukov-2,diffppaamods,diffppaamods-1,diffppaamods-2,diffppaamods-3,diffppaamods-4,diffppaamods-5,diffppaamods-6,epos}.

In the present analysis,  studies of one-dimensional BEC effects in $pp$ collisions at centre-of-mass energies of 0.9 and 
7~\TeV,  using the ATLAS detector~\ct{atlasdet} at the Large Hadron Collider 
(LHC), are presented. At the LHC, BEC have been studied by the CMS \ct{cms-bec1st,cms-JHEP} and ALICE \ct{alice-bec900,alice-bec3d} experiments.
 In   the   analysis   reported here,  the  studies    are extended   to the   region of high-multiplicities   available   thanks 
 to the   high multiplicity track trigger. 
The results are compared to  measurements at the same or  lower energies.

\section{Analysis}
\subsection{Two-particle correlation function}
\label{2partfunct}

Bose--Einstein correlations are measured in terms of a  two-particle correlation function, 
\begin{equation}
 \label{eq_1}
  C_{2} (p_1, p_2) = \frac{\rho (p_1, p_2) }{\rho_0 (p_1, p_2)}\, ,
\end{equation}   
 where $p_1$ and $p_2$ are the four-momenta of two identical bosons  in the 
event,  $\rho$ is the two-particle 
density function, and   
$\rho_0$  is a two-particle  
density function (known as the reference function)  specially constructed to exclude BEC  effects.
The densities $\rho$ and $\rho_0$ are normalized to unity, i.e.\    they are the probability density functions.

In order to compare with data over the widest possible range of 
centre-of-mass  
energies and system sizes, 
the density function 
 is parameterized in terms of
the Lorentz-invariant four-momentum difference squared, $Q^2$, of the  two particles,
 \begin{equation}
 \label{eq_2}
  Q^2 = - (p_1 - p_2)^2 \,  .
 \end{equation}      

The BEC effect is usually described  by  a function with two parameters:  the effective  radius  
parameter
 $R$ and the  strength parameter $\lambda$ \ct{lambdaintro}, where the 
latter   is also called the  incoherence or chaoticity parameter. 
A typical functional form is
\begin{equation}
 \label{eq_3}
  C_{2}(Q) = \frac{\rho (Q) }{\rho_0 (Q)} = C_{0} [1 + 
 \Omega(\lambda, Q R)] (1 + \varepsilon Q) \, .
\end{equation}  
In a simplified scheme 
for fully coherent emission of identical bosons, $\lambda =0$,  while  for incoherent (chaotic) emission, $\lambda =1$. 
The  $QR$ dependence comes from
the Fourier transform of the
 distribution of the  space-time points of boson 
 emission. 
Several different functional
forms have been proposed for  $\Omega(\lambda,QR)$. 
Those used in this paper are described in Sect.\ \ref{sect:params}.  
 The fitted parameter $\varepsilon$ 
 takes into account long-distance correlations   not fully
  removed from $\rho_0$.
Finally, $C_{0}$  is a normalization constant, typically chosen such that 
$C_2 (Q)$ is unity for large $Q$.
In this paper, the density function $\rho$ is calculated for like-sign charged-particle pairs, 
with both the $++$ and $--$ combinations  included,  $\rho (Q) \equiv \rho (++, --)$.
All particles are treated as charged pions  and no particle identification is attempted. 
The purity of the analysis sample in terms of identical boson pairs is estimated from MC to be about 70\% 
(where about 69\% are $\pi^\pm \pi^\pm$  and about 1\% are $K^\pm K^\pm$). 
The effect of the purity is absorbed in the strength parameter $\lambda$, 
while the results of the analysis on the effective radius parameter $R$ were found to be not affected.

\subsection{Coulomb correction 
\label{Coulomb_correction}
}

The  long-range Coulomb force causes  a  momentum shift  between the like-sign and unlike-sign pairs of  particles. 
The density distributions are corrected for this effect by applying the Gamow penetration 
factor per track pair with a weight $1/G(Q)$
\ct{coulomb,coulomb-0,coulomb-1}  (for review see Ref.\ \ct{HI-revs-wied})
\begin{equation}
 \label{eq_10}
  \rho_{\rm corr} (Q) = {\rho (Q) \over G(Q)} \,  , 
\end{equation} 
where  the Gamow factor $G(Q)$ is given by
\begin{equation}
 \label{eq_11}
  G(Q) = \frac{2\pi\zeta}{{\mathrm e}^{2\pi\zeta}-1}
\end{equation} 
 with the dimensionless  parameter $\zeta$ defined as 
 \begin{equation}
 \label{eq_11a}
  \zeta = \pm\frac{\alpha m}{ Q }\, . 
 \end{equation}
 Here $\alpha$  is the electromagnetic fine-structure constant and $m$ is the  pion
mass. 
The sign of $\zeta$ is positive for like-sign pairs and  negative  for unlike-sign pairs.
The resulting correction on $\rho (Q)$ decreases with increasing $Q$ and at 
$Q=0.03$ \GeV\ it is about 20\%. 
A systematic uncertainty on $G(Q)$ is considered to cover effects
 like the extended size of the emission source and other effects, 
see discussion in  Ref.\ \cite{becrev-weiner,becrev-weiner2}.  
Neither the Coulomb interaction nor the BEC effect are present in the generation 
of MC event samples which are used in the analysis. 
The Coulomb correction is thus not applied to MC events.

\subsection{Reference sample}
 \label{subsect:refsample}

 A good choice of the reference sample is important to allow the  experimental detection of the BEC signal.
 Ideally, $\rho_0  (Q)$ should include all 
momentum correlations
except those arising from BEC.
Thus, several different choices have been studied to construct an 
appropriate reference sample.

Most of the  proposed approaches  use random pairing of particles, such as 
mixing particles from different events (the ``mixed event'' technique 
\ct{mixed}),
 or  choosing them from the  same event but from opposite hemispheres or 
 by rotating the transverse momentum  vector of one of the particles  of the like-sign pair \ct{book}.  
 Although  these mixing techniques reproduce the topology and some properties of 
 the event under consideration and  destroy BEC, they violate  
 energy-momentum conservation. 
 Moreover, there are many possible ways to construct the pairs,
 such as mixing the  particles randomly, or keeping some topological constraints such as the 
event multiplicity, the invariant mass of  the pair or  the rapidity of  the pair.
 All of these introduce additional  biases  in the BEC observables.
For example, it was observed in dedicated MC studies that the single-ratio correlation 
functions $C_2$ using reference samples constructed with the event mixing or 
opposite hemispheres techniques exhibit an increase in the low-$Q$ BEC sensitive region.
This effect is found to be more pronounced with increase of the multiplicity or 
average particle-pair transverse momentum and indicates that these reference 
samples are not suitable.

A natural choice is to use the unlike-sign particle pairs from the same 
 events that are used to form pairs of  like-sign particles, i.e., 
  $\rho_0 (Q) \equiv \rho  (+-)$, called in the following the unlike-charge reference sample.
This sample has the same topology and global properties as the like sign 
sample $\rho (++,--)$, 
but is naturally free of any BEC effect.
Studying the $C_2$ correlation functions on  MC, none of the deficits of the event mixing 
and opposite hemispheres techniques described above were observed.
 However, this sample contains hadron pairs from the decay of resonances 
such as  $\rho,\ \eta,
 \eta^{\prime},\ \omega,\ \phi,\   K^{*}$, which are not present in the 
like-sign  combinations. 
These contribute to the low-$Q$ region and can give a spurious BEC signature with a large effective 
radius of the source
\ct{resonances,Gyulassy:1988yr,Padula:1988uf,resonances-1,resonances-2,resonances-3,resonances-4}. 

 In this paper, the unlike-charge reference sample is used. 
 To account for the effects of resonances, the two-particle correlation 
function $C_2(Q)$ is corrected using Monte Carlo simulation without BEC 
effects via a double-ratio $R_2(Q)$ defined as
\begin{equation}
 \label{eq_4}
   R_{2}(Q) = {C_{2}(Q) \over C^{\mathrm MC}_{2}(Q)} = \frac{ \rho {(++,--)} }{ 
\rho {(+-)}} \Bigg/ \frac{ \rho^{\mathrm MC} {(++,--)} }{ \rho^{\mathrm MC} {(+-)}}\, . 
 \end{equation}

\subsection{The parameterizations of BEC }
\label{sect:params}

 Various  parameterizations of the $\Omega (\lambda, Q R)$ function  can be found in  the 
literature,  each  assuming a   different shape for the   particle-emitting source.
 In  the  studies   presented here,   the data are analysed  using   the following   parameterizations: 
 \begin{itemize} 
 \item 
the Goldhaber  parameterization \ct{gglp,gglp2} of a   static Gaussian source  in the plane-wave approach, 
 \begin{equation}
 \label{eq_5}
\Omega=    \lambda \cdot  \exp{\left(- R^{2} Q^{2}\right)} \, ,
 \end{equation}
which assumes a spherical shape with a radial Gaussian distribution of the  
emitter;
\item  
the exponential parameterization  of a static source 
 \begin{equation} 
 \label{eq_6}
  \Omega =  \lambda \cdot \exp{\left(- R Q \right) } \, ,
 \end{equation}
which  assumes a radial Lorentzian distribution of the source.
This parameterization provides a better  description of   the data at 
small $Q$ values, as 
  discussed 
 in  \ct{book}.
 \end{itemize}
The first moment of the  $\Omega (Q R)$ 
distribution  corresponds to  $1/R$  for the exponential 
form  and to $1/(R\sqrt{\pi})$ for the Gaussian form.
To compare the values of the  radius  
parameters
 obtained from the two  functions, the $R$ value of the Gaussian should be compared to  
$R/\sqrt{\pi}$ of the exponential form. 

\section{Experimental details}
\label{expdetails}

\subsection{The ATLAS detector}
\label{atlasdetector}

The ATLAS detector~\ct{atlasdet}  is a multi-purpose particle physics experiment
operating at  one of the beam  interaction points of the LHC.
 The detector  covers almost the whole solid angle around the collision  point with 
layers of tracking detectors, calorimeters and muon chambers. 
It is designed to study a wide range of physics topics at LHC energies. 
For the measurements presented in this paper, the tracking devices and the 
trigger system are of particular importance.

The   innermost part of the   ATLAS  detector is   the inner detector (ID),  
 which  has full coverage in $\phi$ and covers the pseudorapidity range 
$|\eta|<2.5$.\footnote{ATLAS uses a right-handed coordinate system with its origin at the nominal interaction point (IP) 
in the centre of the detector and the $z$-axis along the beam pipe. 
The $x$-axis points from the IP to the centre of the LHC ring, and the $y$-axis points upward. 
Cylindrical coordinates $(r,\phi)$ are used in the transverse plane, $\phi$ being the azimuthal angle around the beam pipe. 
The pseudorapidity is defined in terms of the polar angle $\theta$ as $\eta=-\ln\tan(\theta/2)$.} 
 It consists of a silicon pixel detector (Pixel), a silicon  microstrip detector (SCT) and a transition radiation tracker (TRT).  
These detectors
 are immersed in a $\rm 2\, T$ 
 solenoidal magnetic field. 
 The Pixel, SCT, and TRT detectors have typical  position resolutions 
 of   10, 17 and 130~$\mu $m for the $r$-$\phi$ coordinate, respectively. 
 In the case  of the Pixel and SCT,   the resolutions are     
115 and 580~$\mu$m, respectively, for the second measured coordinate.  
A track from a charged particle traversing the full radial extent
 of the ID 
would typically have
 three Pixel hits, eight 
 or more SCT hits and more than 30 TRT  hits.

 The ATLAS detector has a three-level trigger system: Level~1~(L1), Level~2 (L2) and Event~Filter~(EF). 
For this measurement, the trigger relies on the L1 signals from the Beam Pickup Timing devices (BPTX) and the Minimum-Bias (MB) Trigger Scintillators (MBTS). 
The BPTX are   composed of electrostatic button pick-up detectors attached to the beam  pipe 
and  located 175~m  from the centre of the ATLAS detector in both directions along the beam pipe.
The MBTS are mounted at each end of the detector in front of the liquid-argon end-cap 
calorimeter cryostats at $z = \pm 3.56$~m. 
They are segmented into eight sectors in azimuth and two rings in pseudorapidity 
($2.09 < |\eta|~<~2.82$ and $2.82 < |\eta|~<~3.84$).  
Data was collected requiring coincidence of BPTX and MBTS signals, where only 
a single hit in the MBTS was required on either side of the detector. 
The efficiency of this trigger was studied with  events collected with a separate 
prescaled L1 BPTX trigger,  filtered by ID requirements at L2 and at EF level in
 order to obtain inelastic interactions and found to be 98\% for two
 selected tracks and 100\% for more than four selected tracks \cite{atlas927,atlas900}.

High-multiplicity track  (HM) events were collected at 7~\TeV\ using a  dedicated high-multiplicity track trigger.
At L1, the collisions were triggered using the summed transverse energy ($\Sigma E_{\rm T}$) in 
all calorimeters,  calibrated at  the electromagnetic  energy scale  \ct{ATLAS-LARG-PUB-2008-002}.
The  high-multiplicity
events were required to have $\Sigma E_{\rm T}>20$~\GeV.   
A high number of hits in the SCT was required at L2, while at the EF 
level at least 124 tracks with $\pT > 400$~\MeV\  were  required to originate from  a single vertex.

 \subsection{Data and Monte Carlo samples}
 \label{dataset}

The study is carried out using the $pp$-collision  
 datasets at  the centre-of-mass  energies $\sqrt{s}=$~0.9  and 7~\TeV\  
that   were used  in  previously  published ATLAS studies of minimum-bias  interactions \ct{atlas927,atlas900}. 

The event and track selection criteria are the same as the ones used for the ATLAS minimum-bias multiplicity analysis
\ct{atlas900} with the same minimum-bias trigger and quality criteria for the track  reconstruction. 
 All events in these datasets are required to have at least one vertex \ct{atl-vertex}, formed
from a minimum of two tracks with $\pT  >$~100~\MeV\ and consistent with the average 
beam spot position within the ATLAS detector (primary vertex) \ct{ATLAS:2010gaa}.
The tracks satisfying the above-mentioned selection criteria are used as the input to determine the
corrected distributions, as described in Sect.~\ref{correction}.
The multiplicity of selected tracks with $\pT > 100$~\MeV\ and
$|\eta|<$~2.5 within an event is denoted by $n_{\rm sel}$.

The contributions from beam--gas collision and from non-collision background (cosmic rays 
and detector noise) were investigated in Ref.\ \ct{atlas927} and found 
to  be negligible. Events with more than one  primary vertex (less than 0.3\% of the sample) 
are rejected in order to prevent a bias from multiple proton--proton interactions (pile-up) 
in the colliding proton bunches.

 The same event selection criteria are applied to  high-multiplicity  events, which are defined to be those with at least 120 selected tracks. 
To estimate the possible  influence  of multiple 
 $pp$ interactions in the 7~\TeV\  high-multiplicity  track trigger 
 data,  the distribution of the distances $\Delta z$ between the  $z$   coordinates  of  primary and  pile-up vertices  
 are studied.  
The study shows that on average there is less than one pile-up track selected in the HM sample, 
which has a negligible influence on the BEC studies.

For the measurements at $\sqrt{s}=0.9$~\TeV, 
about $3.6\times 10^{5}$ events with a total of 
more than $4.5\times 10^{6}$  tracks 
 are after selection,  
and in the case of $\sqrt{s}=7$~\TeV, 
about $10^{7}$ events with   
about $2.1\times 10^{8}$ tracks   overall
 are after selection.
 This corresponds to integrated luminosities of 
$\sim$7~$\mu$b$^{-1}$ 
and  $\sim$190~$\mu$b$^{-1}$  at  0.9~\TeV\  and  7~\TeV, respectively.
 For the    measurements at 7~\TeV\    with the  high-multiplicity track trigger,  
 about $1.8\times10^{4}$  events  with  
 more than $2.7\times 10^{6}$  tracks overall were after selection.
 This corresponds to integrated luminosity of  
 $\sim$12.4~nb$^{-1}$. 

 Large Monte Carlo  samples of   minimum-bias 
and high-multiplicity
events  were generated using the PYTHIA 6.421  Monte Carlo  event  generator \ct{pythia6} 
with the ATLAS MC09 set of optimised parameters (tune) \ct{p6atlas} 
($1.1 \times 10^7$ for $\sqrt{s}=900$ \GeV,  $2.7 \times 10^7$ for $\sqrt{s}=7$ \TeV\ 
and $1.8 \times 10^6$ for $\sqrt{s}=7$ \TeV\  high-multiplicity data) 
with non-diffractive, single-diffractive and double-diffractive processes
included in proportion to the cross sections predicted by the model.
 As discussed in Sec.\ \ref{Coulomb_correction},  no simulation of the BEC effect is implemented in the generator.
This is  the  baseline Monte Carlo generator which  reproduces   single-particle spectra  \ct{atlas927}.
The generated events  were pas\-sed through the ATLAS simulation and reconstruction chain; 
the detector simulation program \ct{atlsim} is based on  GEANT4 \ct{geant4}. 
 Dedicated sets of high-multiplicity  events were also generated.
 
For the  study of systematic  effects, additional Monte  Carlo samples
 were produced using the PHOJET 1.12.1.35 generator \ct{phojet},  PYTHIA 
with the Perugia0  tune  \ct{perugia}; 
and  the  EPOS 1.99\_v2965 generator \ct{epos}  for the  high-multiplici\-ty  analysis. 
The PHOJET  program uses the Dual Parton Model \ct{dpm} for low-$\pT$  physics and 
is interfaced to  PYTHIA  for  the  fragmentation  of partons.
The EPOS generator is based on an implementation of the QCD-inspired Gribov--Regge field theory describing soft and 
hard scattering  simultaneously,   and relies on the  same parton distribution functions as used in PYTHIA.
 The EPOS LHC tune is   used with  parameters optimised to describe the LHC minimum-bias  data 
 \ct{atl-et}.    

The high-multiplicity PYTHIA MC09 and EPOS samples, each are about two magnitudes larger than the data sample. 
The $C_2(Q)$ single-ratio  correlation functions in MC reproduce data well for $Q> 0.5$ GeV. 
In the region $Q<0.5$ GeV, the BEC effect is clearly seen in the data $C_2(Q)$ 
correlation function while no such effect is seen in the MC as expected, 
since  no BEC present in MC.

 \subsection{Data correction procedure}
 \label{correction}

Following the procedure applied in the  
previous ATLAS minimum-bias measurements \ct{atlas927,atlas900}, 
each track is assigned a weight which corrects for the track reconstruction 
efficiency, for the fraction of secondary particles, for the  fraction of the primary
 particles\footnote{In the Monte Carlo simulations, 
primary charged particles are defined as  charged particles with a mean 
lifetime $\tau > 0.3 \times 10^{-10}$~s either directly produced in $pp$ 
collisions or from the subsequent decay of particles with a shorter  lifetime.}
 outside the   kinematic range and for the   fraction of fake 
tracks.\footnote{Fake tracks are  tracks constructed from tracker 
noise and/or  hits which   are not produced by a single-particle.}
 In addition,  the effect of  events lost due to trigger and 
vertex reconstruction inefficiencies is corrected for  using   an
 event-by-event weight applied to pairs of  particles in the $Q$ 
 distribution. 
The efficiency of the high-multiplicity track trigger has been studied in data as a function of the number 
of reconstructed tracks and is found to be 5\% for 120 selected tracks and to reach 
 a plateau at  100\% once 150 tracks are selected. 
The measured trigger inefficiency is used to correct the experimental 
distributions and  is  found 
to have negligible  impact on the extraction of the BEC parameters discussed in Sec.\  \ref{results}.

The multiplicity distributions are corrected to the particle level using 
an iterative method that follows the Bayesi\-an approach 
\ct{unfold-agostini} as it is 
described in  Refs.\ \ct{atlas927,atlas900}.  
 An unfolding  matrix reflecting the probability of reconstructing $n_{\rm 
sel}$ charged tracks in an event with 
generated  charged-particle  multiplicity $n_{\rm ch}$ is populated using Monte Carlo simulation and applied to the data. 
The unfolding matrix is built using the  ATLAS  MC09   PYTHIA  tune \ct{p6atlas}.
The unfolding procedure converges after the fifth iteration. 
It is found that the corrected multiplicity distribution agrees well with the published result \ct{atlas927,atlas900}.
The unfolding procedure of the 7 \TeV\ high-multiplicity data follows the same technique and 
unfolding matrix used in the previous analysis of minimum-bias data in 
Ref.\ \ct{atlas927}, 
 restricted to the region of high charged particle multiplicity specific 
 to this analysis, and  convolved
 with a normalised Gaussian distribution to account for 
the experimental 
resolution on the number of selected tracks.
It is found that a number of 120 selected tracks at detector level, $n_{\rm sel}$, 
corresponds to about 150 charged particle, $n_{\rm ch}$, at particle level.
Momentum distributions are unfolded in a similar way.

For all distributions, closure tests are carried out using Monte Carlo
samples corrected according to the same procedure as used in the data. 
 The difference obtained between the reweighted distributions and those at the particle  level is due to 
 tracking effects such as a smaller  reconstruction efficiency  for pairs of tracks with very small opening angle. 
These effects are small for correlation functions constructed using data, 
typically 1--3\%, and are included in the systematic uncertainty.
In the case of the unfolded $Q$ distributions, the  data are corrected for the bias from secondary tracks using 
Monte Carlo simulation and the corresponding systematic  uncertainty is obtained 
by variation of the  amount of material in the inner detector 
 by  $\pm 10$\%.

\section{Systematic uncertainties  
}

The systematic uncertainties of the inclusive  fit parameters, $R$ and $\lambda$, of  the  exponential model are summarized in Table~\ref{tab:systr2}.
The following contributions to the systematic uncertainties on the fitted parameters are considered. 

 The systematic uncertainties resulting from  the  track  reconstruction efficiency, 
which are  parameterized in bins of $\pT$ and $\eta$, were determined   in earlier   analyses  \ct{atlas927,atlas900}. 
These cause uncertainties in
the track weights  of particle pairs  in the $Q$ distributions  entering the correlation functions.
 
The effects of track splitting and merging are
sizeable only  for very  low $Q$ values  (smaller than 5 \MeV), 
and are  found to be negligible  for the measurements with $Q \ge 20$ \MeV.

 The leading source of systematic uncertainty   is due to differences in 
the Monte Carlo generators used to calculate the $R_2$ correlation 
function from the 
$C_2$ correlation function. 
 The corresponding contribution  to the systematic   uncertainty is estimated  as the root-mean-squared (RMS) spread 
of the  results obtained for the different Monte Carlo   datasets.
 The statistical uncertainties arising from the Monte Carlo datasets   are negligibly small. 

 The systematic  uncertainty 
due to Coulomb corrections  is estimated by varying  the corrections  by $\pm 20$\%.

The influence of the fit range is estimated  by changing 
the  upper bound of the $Q$ range from the nominal  2~\GeV:  decreasing it to
 1.5~\GeV\  and increasing  it  up to 2.5~\GeV. 
The   latter  better estimates the  uncertainty due the  long-range correlations. 
This contribution is taken
into account by the value of $\varepsilon$, the parameter in
 the linear term of Eq.~(\ref{eq_3}) describing the long-range  correlations.

Other effects contributing to the systematic uncertainties  are the lowest
value of $Q$ for the fit,  the bin size  and  exclusion of the  interval  $0.5\le Q \le 0.9$~\GeV\
due to the overestimate  of the $\rho$ meson contribution in the Monte Carlo simulations, 
as discussed in the following Sec.\ \ref{twoparticle}. 
These uncertainties are estimated by varying the lowest $Q$ value in the fit  by $\pm 10$~\MeV, 
by changing the bin size by $\pm 10$~\MeV, and by broadening  the excluded interval 
by 100~\MeV\ on both sides.

The background of photon 
conversions into $e^+e^-$ pairs was studied and found to be negligible.

To test the effect of treating all charged particles as pions, the double-ratio correlation 
functions $R_2$ are also obtained using only identical particles in the Monte Carlo sample to compute the correction. 
The resulting BEC parameters fitted to the $R_2$ functions defined this way show 
negligible differences to the nominal result and no further systematic 
uncertainties are assigned.

 Finally, the systematic uncertainties are combined by adding them in quadrature and
 the resulting values are given in the bottom row of  Table   \ref{tab:systr2}. 
 
The same sources  of uncertainty are considered for the differential measurements in 
$n_{\rm ch}$ and the average transverse momentum $\kT$ of a pair, 
and their impact on the fit parameters is found to  be similar in size.

\begin{table*}[t!]
\caption{
Systematic uncertainties on $\lambda$ and $R$ for the exponential fit 
of the two-particle  double-ratio correlation function  $R_2 (Q)$ in the full kinematic region at $\sqrt{s}=$0.9 and 7~\TeV\ 
for minimum-bias and high-multiplicity 
events.
}
\label{tab:systr2}
\begin{center}
 \begin{tabular}{lcccccccc}
\hline
\hline
& \multicolumn{2}{c}{0.9~\TeV} && 
\multicolumn{2}{c}{7~\TeV} && 
\multicolumn{2}{c}{7~\TeV\ (HM)} \\  
 
 \hline 
Source       &   $\lambda$ &  $R$ & &  $\lambda$ &  $R$ & &  $\lambda$ & $R$ 
 \\
\hline
Track reconstruction efficiency 		&  0.6\% 	& 0.7\% 	&&   0.3\%  	&  0.2\%  	&&  1.3\%  	&  0.3\%    \\ 
Track splitting and  merging      &  
 \multicolumn{2}{c}{negligible} && \multicolumn{2}{c}{negligible} && \multicolumn{2}{c}{negligible} \\  
Monte Carlo samples 			&  14.5\% 	&   12.9\%   &&   7.6\%   	& 10.4\%   	&&  5.1\%  	&   8.4\%  \\
Coulomb  correction 			&   2.6\% 	&   0.1\% 	&&   5.5\%  	& 0.1\%  	&&  3.7\%  	&  0.5\%    \\
Fitted range  of $Q$      			&     1.0\% 	& 1.6\% 	&& 1.6\% 	& 2.2\%  	&& 5.5\%  	&  6.0\%    \\ 
Starting value  of $Q$          		&   0.4\% 	& 0.3\% 	&& 0.9\% 	& 0.6\% 	&&  0.5\%  	&  0.3\%    \\
Bin size         					&  0.2\% 	& 0.2\% 	&& 0.9\% 	& 0.5\% 	&& 4.1\%  	&  3.4\%    \\
Exclusion interval 				& 0.2\% 	& 0.2\% 	&&  1\% 	& 0.6\% 	&& 0.7\%  	&  1.1\%    \\ 
\hline
Total 						&  14.8\% 	&   13.0\% 	&&    9.6\%  &     10.7\% 	&&  9.4\%  	&  10.9\%   \\ 
\hline
\hline
\end{tabular}
\end{center}
\end{table*}

\section{Results}
\label{results}
 
\subsection{Two-particle correlations}
\label{twoparticle}

\begin{figure*}[t!]
\begin{minipage}[h]{0.50\textwidth}
\center{\includegraphics[width=1\linewidth]
{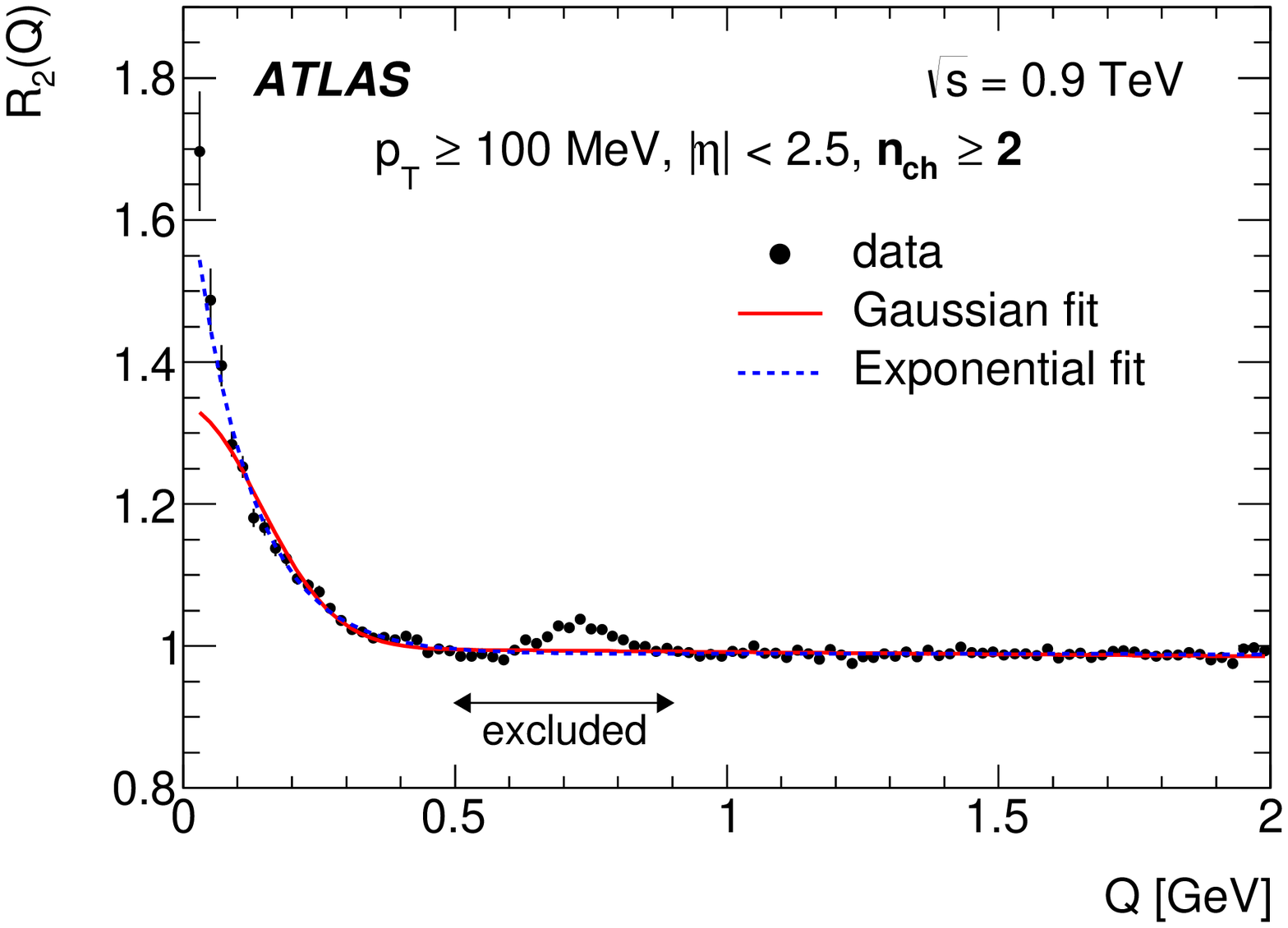}} 
(a) \\
\end{minipage}
\hfill
\begin{minipage}[h]{0.50\textwidth}
\center{\includegraphics[width=1\linewidth,height=0.75\textwidth]
{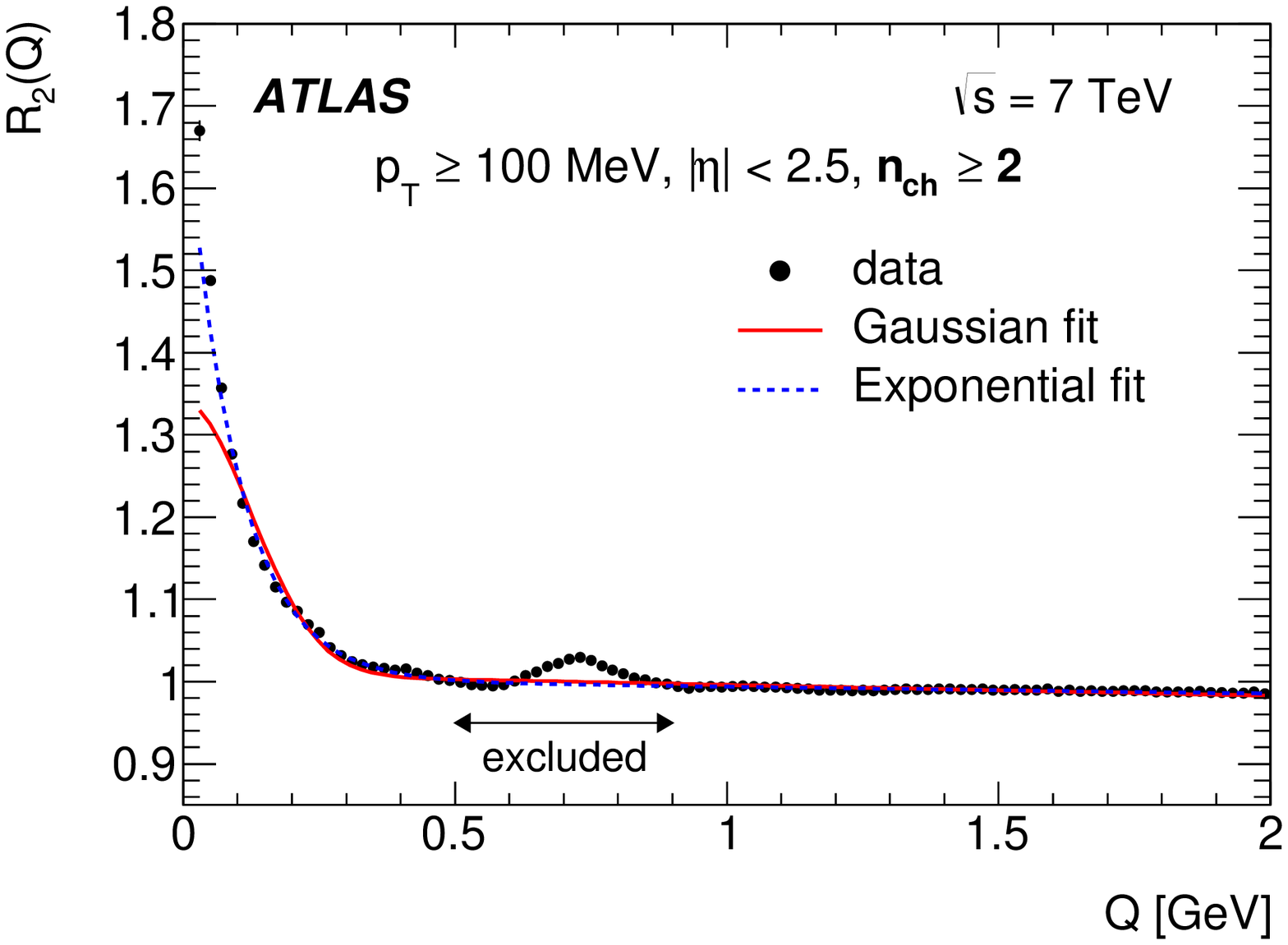}} 
(b)\\
\end{minipage}
 \vfill
 \begin{center}
 \begin{minipage}[h]{0.50\textwidth}
 \center{\includegraphics[width=1\linewidth]
{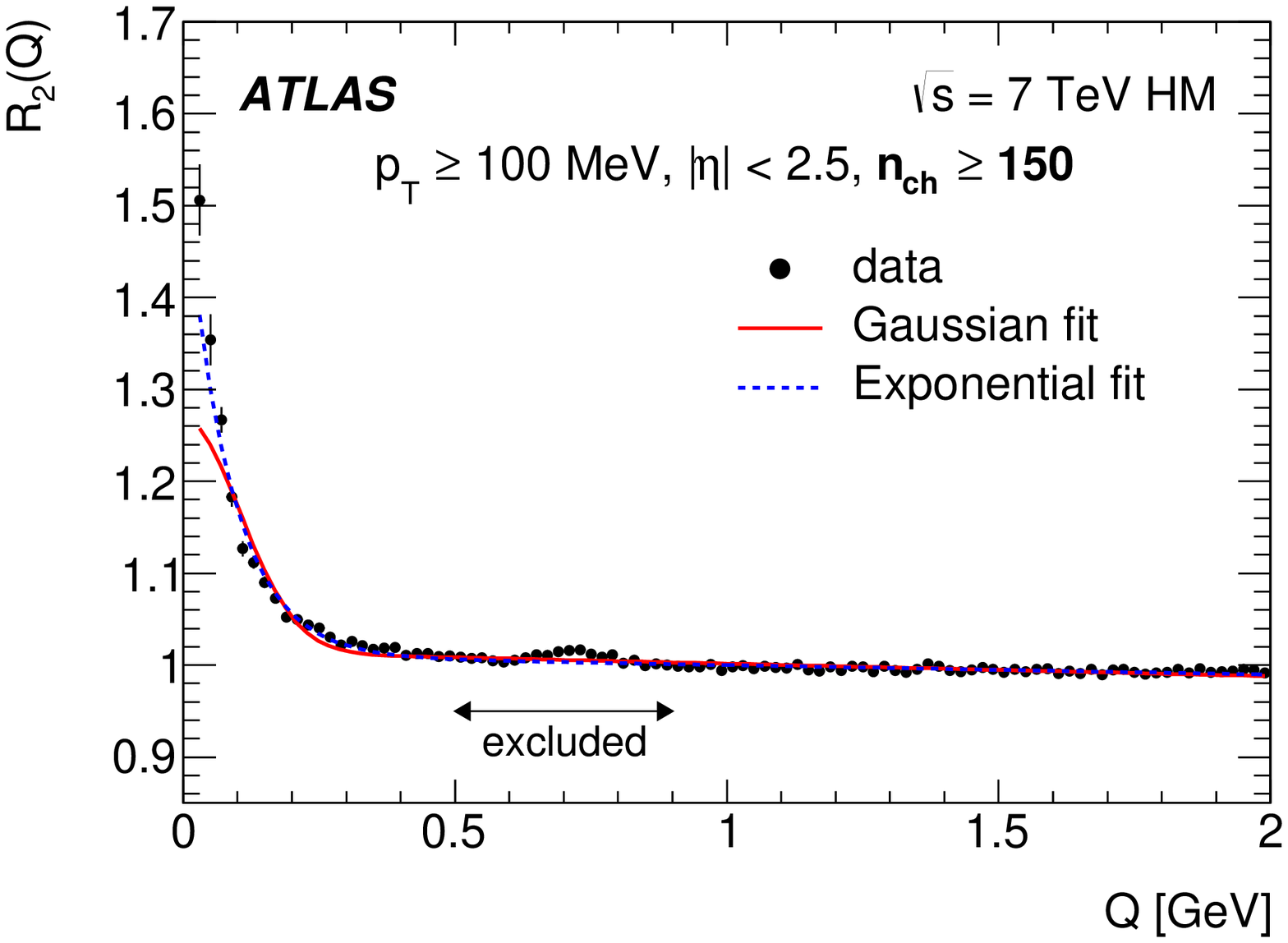}}  
(c) \\ 
\end{minipage}
\end{center}
\caption{
The two-particle double-ratio correlation function $R_2 (Q)$ for  charged particles in $pp$ collisions  at (a) $\sqrt{s}=$0.9~\TeV, (b) 7~\TeV\   
and (c) 7~\TeV\  high-multiplicity   events. 
 The lines show the Gaussian and exponential fits as described in the legend.  
The  region excluded from the fits  is  indicated. 
The  error bars represent the statistical uncertainties.
}
\label{Fig_1}
\end{figure*}

In Fig.\  \ref{Fig_1} the double-ratio $R_2(Q)$ distributions,  measured for 
 0.9  and 7~\TeV,  are compared with 
 Gaussian and exponential fitting functions, Eqs.\  (\ref{eq_5}) and 
(\ref{eq_6}). 
  The fits are performed in the $Q$ range   0.02~\GeV\ to 2~\GeV\   and with  a bin   width of 0.02~\GeV. 
 The upper $Q$   limit is chosen to  
be far away from the low-$Q$ region, which is sensitive to BEC  effects and resonances.
Around $Q\sim 0.7$~\GeV\, there is a visible bump which is  
due to an overestimate of $\rho  \rightarrow \pi^+ \pi^-$ decays  in the Monte Carlo simulation.
 Therefore the region $0.5 \le Q \le 0.9$~\GeV\  is excluded from the fits.
  As seen in   Fig.\ \ref{Fig_1}, the Gaussian function does not describe 
 the low-$Q$ region while the exponential  function provides a  good description of the data.

The resolution of  the  $Q$ variable  is  better than  10~\MeV\  for  the  
region  most sensitive to BEC effect, $Q < 0.4$~\GeV.
The $Q$ resolution is included in the fit of $R_2$ by convolving the fitting function with a 
Gaussian detector resolution function.
The change in the fit results from those with no convolution applied is found to be negligible.

In the process of fitting $R_2(Q)$ with the exponential function, large $\chi^2$ 
values are observed, in particular for the 7~\TeV\ sample where statistical uncertainties 
 on the fitted data points are below 
 2--4\%. 
These  large $\chi^2$ values can be traced back to a small number of 
individual points or small cluster of points. 
The removal of these points does not change the results of the fit while 
the $\chi^2$ substantially improves. 
In the analysis of the 7~\TeV\ data, for most of the considered cases, 
the expected statistical uncertainties are small compared to the 
systematic ones, therefore 
only total uncertainties on the fitted parameters are given. 
 The latter include the statistical uncertainties rescaled by 
 $\sqrt{\chi^2/\mathrm{ndf}}$ 
 \ct{pdg-chi2}. 
For consistency, the same treatment is applied  to the 0.9~\TeV\ analysis 
where the statistical uncertainties are of the  same order of magnitude as 
the systematic ones.

The results of BEC parameters for 
exponential 
fits of the two-particle  double-ratio  correlation function  $R_2 (Q)$  for events  with 
the  unlike-charge reference sample   are
\\
 \vspace*{-0.4cm}
{
\scriptsize 
\begin{eqnarray}
\nonumber
& & \lambda =  0.74\pm0.11,\: R  =  (1.83\pm0.25)\ {\rm fm}\:\,
{\rm at}\: \sqrt{s} =  0.9 \TeV\:  {\rm for}\:  n_{\rm ch} \ge 2,\\
\nonumber
& &   \lambda = 0.71\pm0.07,\: R= (2.06\pm0.22)\ {\rm fm}\:
{\rm at}\: \sqrt{s} =  7 \TeV\:  {\rm for}\:  n_{\rm ch} \ge 2,\\ 
\nonumber
& & 
\lambda = 0.52\pm0.06,\: R= (2.36\pm0.30)\ {\rm fm}\:
{\rm at}\: \sqrt{s} =  7 \TeV\:  {\rm for}\:  n_{\rm ch} \ge 150.
\end{eqnarray}
 }
 \vspace*{-0.4cm}

The  values  of the fitted parameters are close to the values obtained 
by the CMS \ct{cms-JHEP} and ALICE \ct{alice-bec900} experiments.

\subsection{Multiplicity dependence}
\label{sect:multiplicity}

\begin{figure*}[t!]
\begin{minipage}[h]{0.50\textwidth}
\center{\includegraphics[width=1\linewidth]
{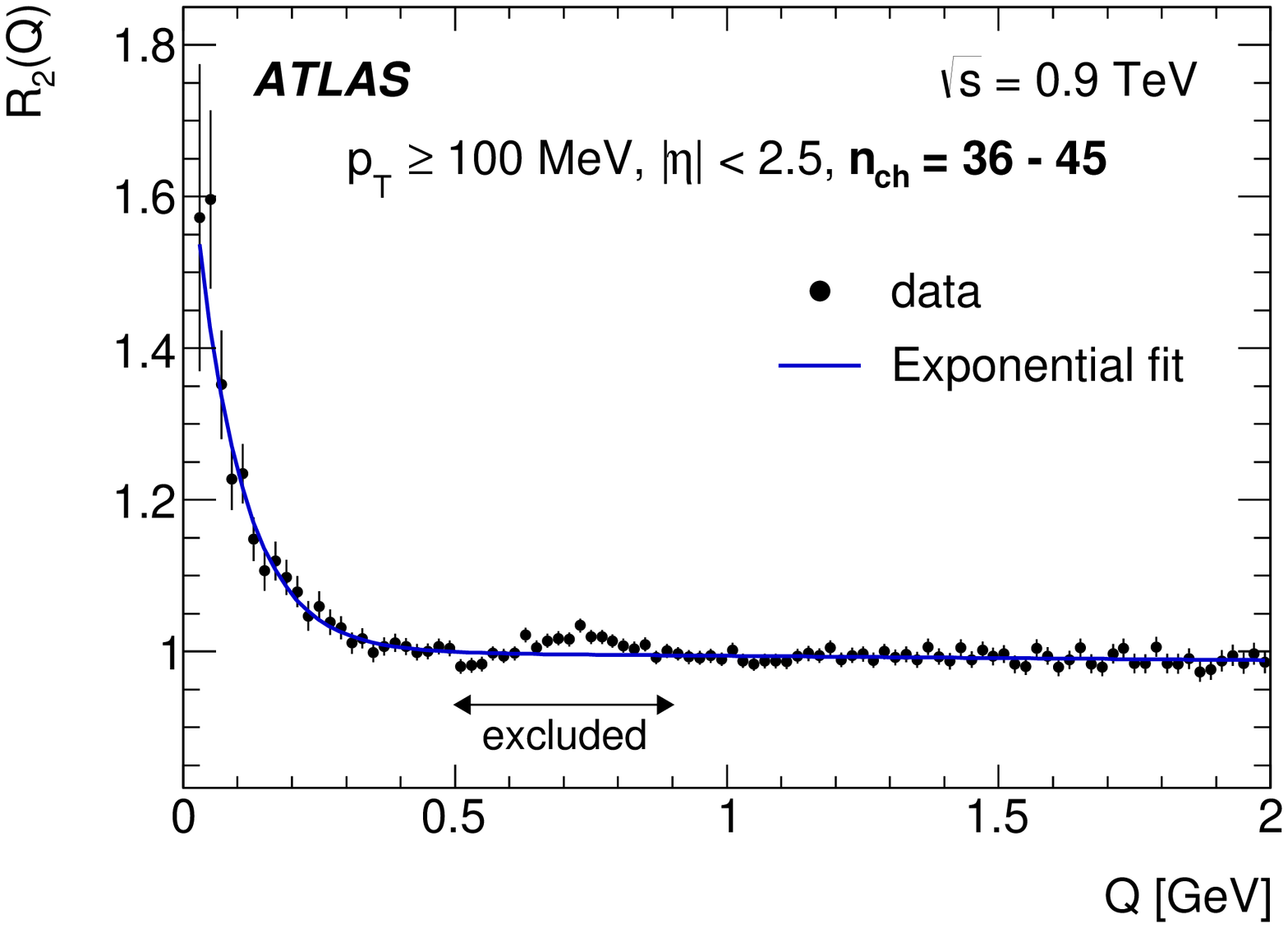}} 
(a) \\[-2mm]
\end{minipage}
\hfill
\begin{minipage}[h]{0.50\textwidth}
\center{\includegraphics[width=1\linewidth,height=0.75\textwidth]
{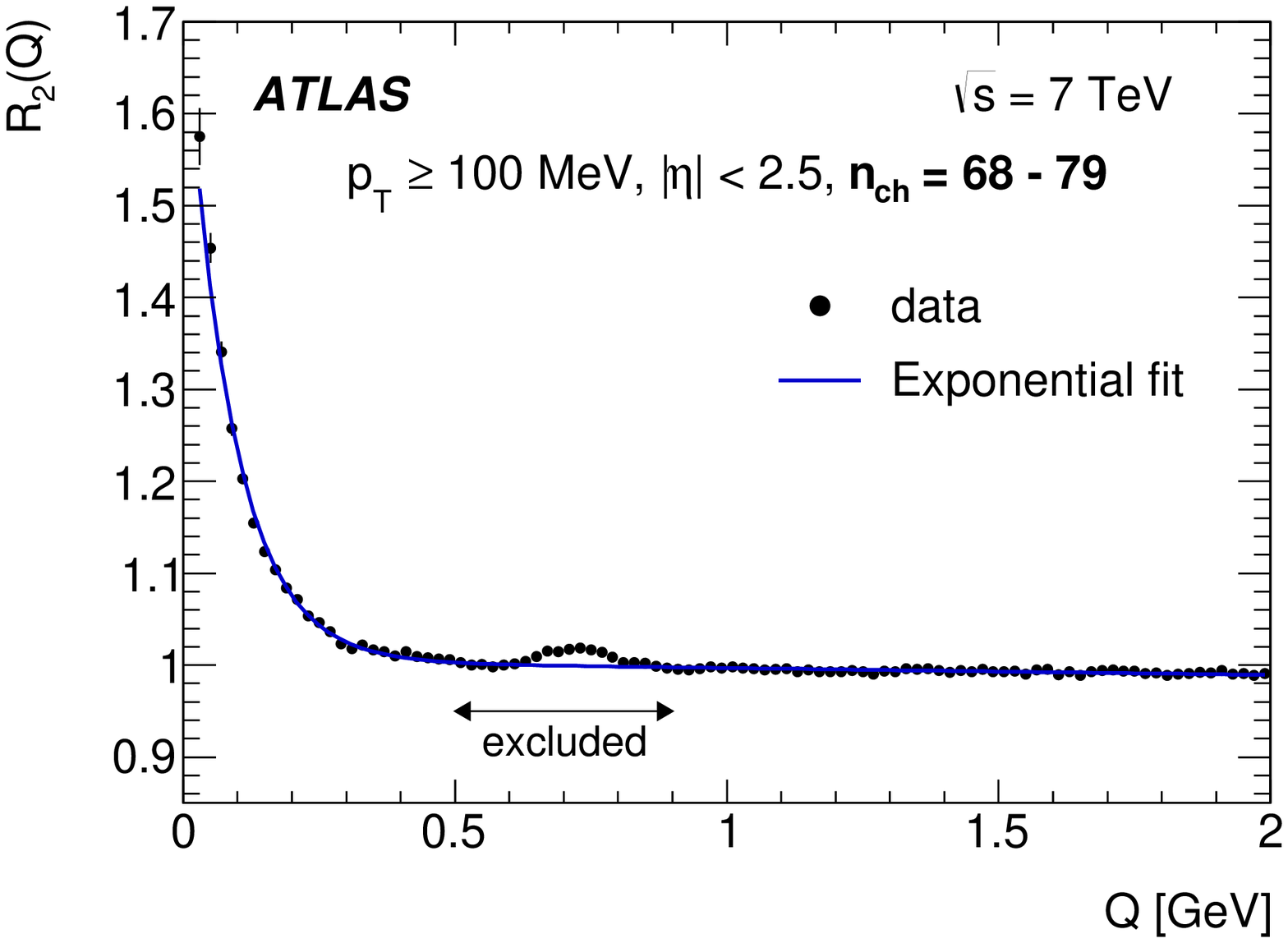}} 
(b)\\[-2mm]
\end{minipage}
 \vfill
 \begin{center}
 \begin{minipage}[h]{0.50\textwidth}
 \center{\includegraphics[width=1\linewidth]
{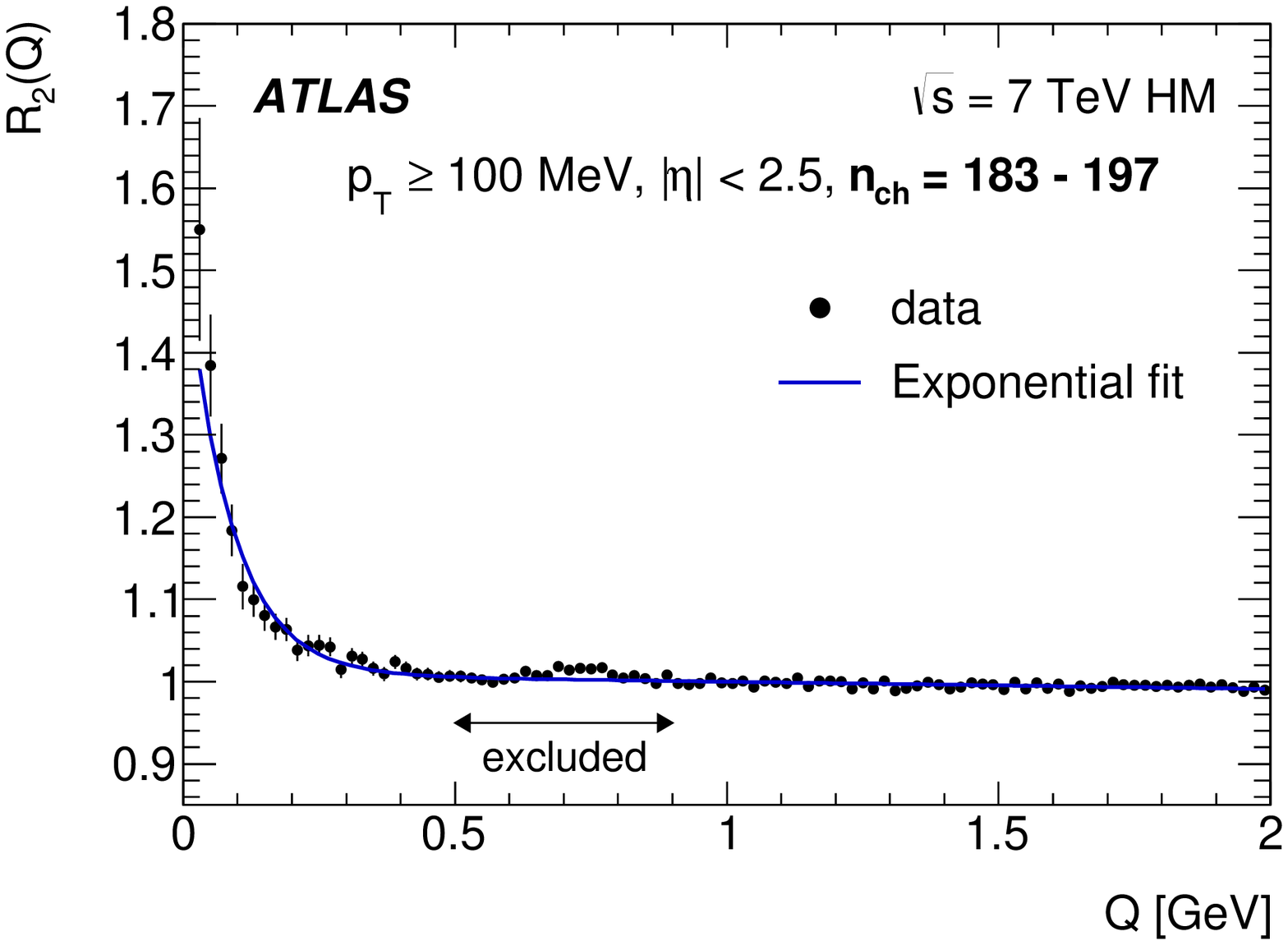}}  
(c) \\[-2mm] 
\end{minipage}
\end{center}
\vspace{-4mm}
\caption{
The two-particle double-ratio correlation function $R_2 (Q)$ for  charged particles in $pp$ collisions  for multiplicity intervals 
(a) $36 \le n_{\rm ch} < 45$ at $\sqrt{s}=$0.9~\TeV, 
(b) $68 \le n_{\rm ch} < 79$ at 7~\TeV\   and 
(c) $183 \le n_{\rm ch} < 197$  at 7~\TeV\   high-multiplicity events. 
 The lines show the  results of the exponential fit.  
The region  excluded from the fits is indicated. 
The  error bars represent the statistical uncertainties.
}
\label{Fig_2_1b}
\end{figure*}

 The $R_2(Q)$ functions defined in  Eq.\   (\ref{eq_4}), 
are shown  for  various multiplicity intervals in   Fig.\ \ref{Fig_2_1b}
for 0.9~\TeV,   7~\TeV\   and 7~\TeV\  high-multiplicity  data.  
 The multiplicity intervals  are chosen so as to be similarly populated and  
comparable to those used by other LHC experiments \ct{cms-bec1st,cms-JHEP,alice-bec900,alice-bec3d}.    
Only the    exponential fit is shown.  As in the fit procedure  for the inclusive case, the detector $Q$ resolution is 
included in the fits.

Within  the multiplicity studies,  the BEC parameters are also measured by  excluding 
the low-multiplicity events,  $n_{\rm ch}< 8$, expected to be contaminated by  diffractive physics \ct{atlas927}. 
  No  noticeable changes in the strength  
and  radius  
parameters  
  for $n_{\rm ch} \geq 8$  are observed  compared  to  the full  multiplicity  range for $n_{\rm ch} \geq 2$.

\begin{figure*}[t!]
\begin{center}
\begin{minipage}[h]{0.80\textwidth}
\center{
\includegraphics[width=1\linewidth]
{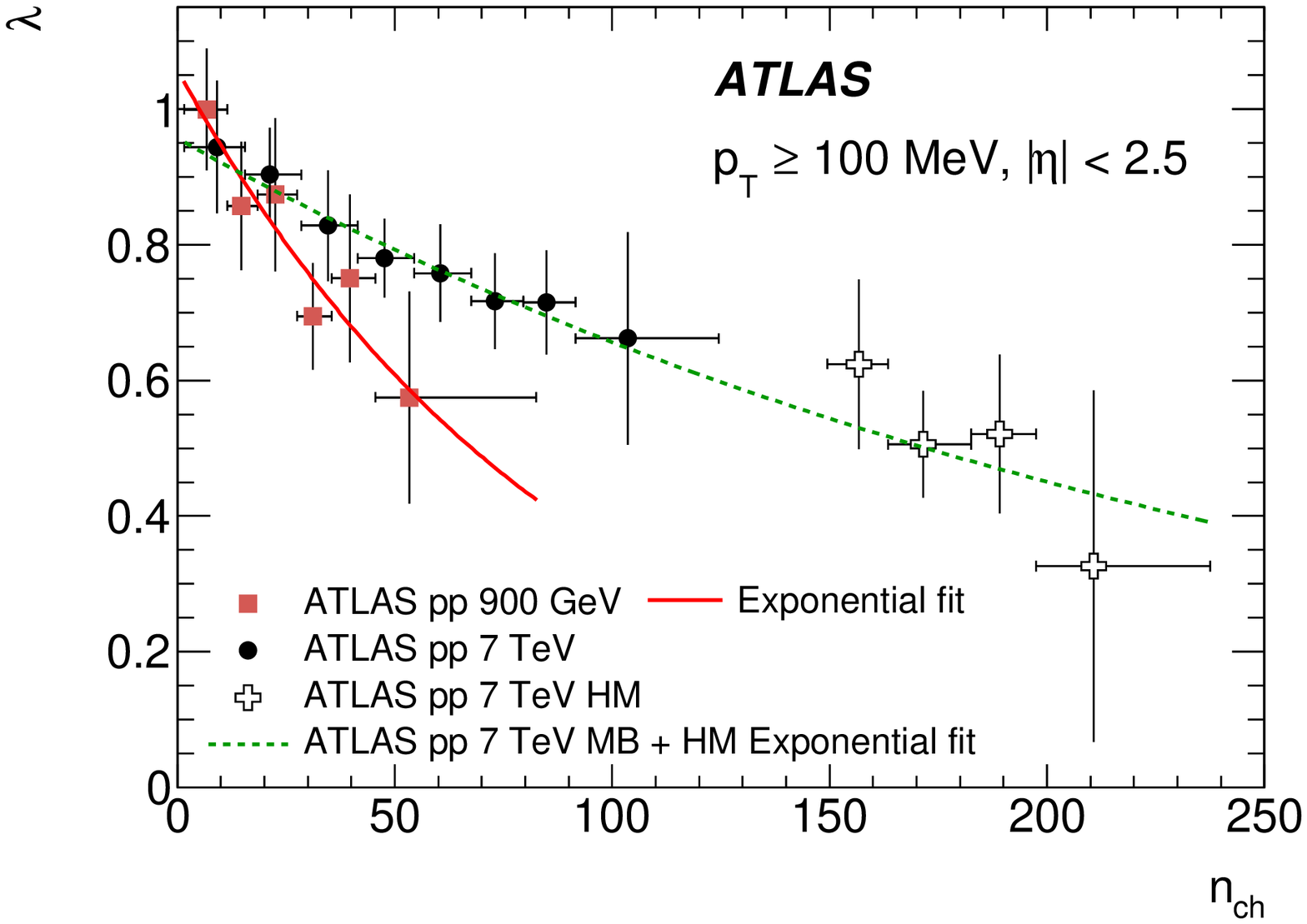} 
} 
(a) \\
\end{minipage}
\end{center}
\vfill
\begin{center}
\begin{minipage}[h]{0.80\textwidth} 
\center{
\includegraphics[width=1.0\linewidth]
{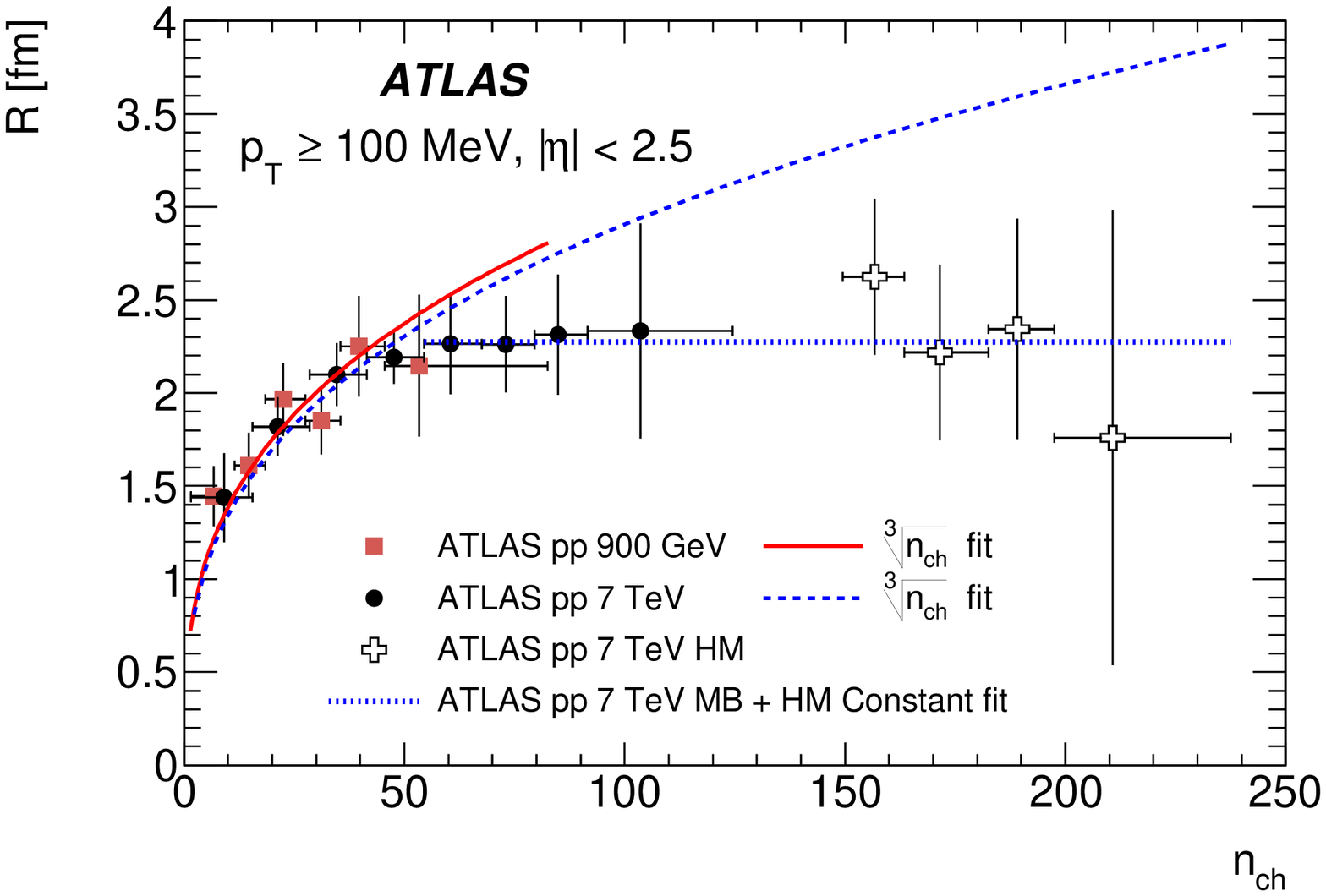}}
(b) \\
\end{minipage}
\end{center}
\caption{
Multiplicity, $n_\mathrm{ch}$, dependence of the 
 parameters (a)  $\lambda$  and (b) $R$  obtained from the exponential fit to  the two-particle double-ratio  correlation functions $R_2 (Q)$ at $\sqrt{s}=$0.9  and 7~\TeV.
The  solid  and dashed curves  are the results of (a) the exponential  and 
(b) $\sqrt[3]{n_{\rm ch} }$ for  $n_{\rm ch} < 55$ fits.
The dotted line in (b) is a result of a constant fit to minimum-bias and high-multiplicity events
data at  7~\TeV\ for $n_{\rm ch} \geq 55$. 
 The error bars represent the quadratic sum of the statistical and systematic uncertainties.
}
\label{Fig_3_1}
\end{figure*}

The multiplicity dependence of the $\lambda$ and   $R$ parameters is shown   in Fig.\   \ref{Fig_3_1}.  
The $\lambda$ parameter decreases with multiplicity, faster for 0.9~\TeV\ than for 7~\TeV\ interactions.
The decrease of  the $\lambda$ parameter with $n_{\rm ch}$ is found to be well fitted 
 with the  exponential function 
$\lambda (n_{\rm ch}) = \gamma\, {\mathrm e}^{- \delta n_{\rm ch}}$.
The fit parameter values are  presented in  Table  \ref{tab:parameters} for 
0.9~\TeV\ and for the combined nominal and high-multiplicity 7~\TeV\  data.

The  $R$ parameter  increases with  multiplicity  up to about $n_{\rm ch} 
\simeq 50$ independently of the center of mass energy. For higher 
multiplicities, the measured $R$ parameter is observed to be independent of multiplicity.
For $n_{\rm ch} \leq 82$ at 0.9~\TeV\ and $n_{\rm ch} < 55$ at 7~\TeV\   
the $n_{\rm ch}$ dependence of $R$  is fitted with the function 
$R (n_{\rm ch}) = \alpha \sqrt[3]{n_{\rm ch} }$,   
similar to that  used  in heavy-ion studies \ct{Lisa:2005dd,alice-bec3d}. 
 The results of the fit are presented in  Table \ref{tab:parameters}  and 
are close to  the  CMS  results \ct{cms-JHEP}.
The fit parameters do not change significantly within uncertainties if data points 
with $n_{\rm ch}>55$  are included in the fit, while the quality of the fit 
significantly degrades.
Therefore the fit is limited to the data points with $n_{\rm ch}\le 55$.
The $n_{\rm ch}$ dependence of $R$ at 7~\TeV\   is  fitted with  a  constant  
$R (n_{\rm ch}) =  \beta$ for  $n_{\rm ch}>  55$;    the resulting value  is given in  Table \ref{tab:parameters}. 
Qualitatively CMS \ct{cms-JHEP}  and UA1 \ct{ua1-bec-diffE} results 
for the radius parameter  follow the same trend as a function of $n_{\rm ch}$ 
as ATLAS data points up to $n_{\rm ch} \le 55$.
The ATLAS and ALICE\cite{alice-bec900,alice-bec3d} results on the multiplicity dependence 
of the radius parameter cannot be directly compared due to much 
narrower $\eta$ region used by ALICE.

 The observed change  of the fitted parameters with  multiplicity  has  been predicted
 in Refs. \ct{book,bec-mult,bec-mult-1,bec-mult-12,bec-mult-2,bec-mult-22}, and is similar to the one also observed in
$e^+ e^-$  interactions  \ct{opal-BEC-mult},  however the saturation of  $R$ for very high multiplicity  is observed 
for the first  time. 

The saturation of $R$ at high multiplicities is expected in a Pomeron-based model 
\ct{schegelsky,schegelsky-1}
as the consequence of the overlap of colliding protons, with the value of the  radius   
parameter
 at $n_{\rm ch}\approx 70$  close to the one obtained in the present studies. 
 However, the same model predicts that above $n_{\rm ch}\approx 70$, $R$ 
will decrease with multiplicity, returning to its low-multiplicity value 
 which is not supported by the data.

\begin{table*}[t!]
\caption{
Results of fitting the multiplicity, $n_\mathrm{ch}$, and the transverse momentum of the pair, $k_\mathrm{T}$, dependence of the BEC parameters $R$ and $\lambda$ with different functional forms and for different data samples.
 The error 
represent the quadratic sum of the statistical and systematic uncertainties.
}
\label{tab:parameters}
\vspace{-3mm}
\begin{center}
 \begin{tabular}{cclll}
\hline
\hline
BEC 	
& Fit 
& \multicolumn{1}{c}{0.9~\TeV} 
& \multicolumn{2}{c}{7~\TeV} 
\\
 \cline{4-5} 
param.\
&function
& 
& \multicolumn{1}{c}{Minimum-bias events} 
& \multicolumn{1}{c}{High-multiplicity 
events}
\\

 \hline

$R(n_{\rm ch})$ 
& $\alpha \sqrt[3]{n_{\rm ch} }$
& $\alpha = 0.64\pm0.07$~fm ($n_{\rm ch} \leq 82$)
& $\alpha = 0.63\pm0.05$~fm ($n_{\rm ch} < 55$)
& \multicolumn{1}{c}{ --- }
\\
 
& $\beta$
&\multicolumn{1}{c}{ --- }
&  \multicolumn{2}{c}{$\mbox{~~~~~~~~} \beta = 2.28\pm0.32$~fm 
($n_{\rm ch} \geq 55$)}
\\
&&&&\\

$\lambda(n_{\rm ch})$ 
& $\gamma\, {\mathrm e}^{- \delta n_{\rm ch}}$
& $\gamma = 1.06\pm0.10$
& \multicolumn{2}{c}{$\gamma =0.96\pm0.07$ \mbox{~~~~~~~~~}}
\\

& 
& $\delta = 0.011\pm0.004$ 
& \multicolumn{2}{c}{$\delta =0.0038\pm0.0008$ \mbox{~~~}}
\\
&&&&\\

$R(\kT)$ 
& $\xi\, {\mathrm e}^{- \kappa \kT}$
& $\xi = 2.64\pm0.33$~fm
& $\xi = 2.88\pm0.27$~fm
& $\xi = 3.39\pm0.54$~fm
\\
 
& 
& $\kappa = 1.48\pm0.67$~\GeV$^{-1}$
& $\kappa = 1.05\pm0.58$~\GeV$^{-1}$    
& $\kappa = 0.92\pm0.73$~\GeV$^{-1}$
\\
&&&&\\

$\lambda(\kT)$ 
& $\mu\, {\mathrm e}^{- \nu \kT}$
& $\mu = 1.20\pm0.18$
& $\mu = 1.12\pm0.10$ 
& $\mu = 0.75\pm0.10$
\\

& 
& $\nu = 2.00\pm0.35$~\GeV$^{-1}$
& $\nu = 1.54\pm0.26$~\GeV$^{-1}$ 
& $\nu = 0.91\pm0.45$~\GeV$^{-1}$ 
\\

\hline
\hline
\end{tabular}
\end{center}
\end{table*}

\subsection{Dependence on the transverse momentum of the particle pair}

 The average transverse momentum  $\kT$ of a particle pair is
 defined as half of the 
 magnitude of the  vector   sum of the two transverse momenta, 
$\kT = |{\bf p}_{{\rm T}, 1} + {\bf p}_{{\rm T},  2}|/2$.  
 The study is performed in the 
 $\kT$ intervals which are chosen in a way 
 to be  similarly populated  and, as for the  multiplicity bins,  to be similar  to the 
intervals used by other  LHC  experiments \ct{cms-bec1st,cms-JHEP,alice-bec900,alice-bec3d}.

\begin{figure*}[t!]
\begin{minipage}[h]{0.50\textwidth}
\center{\includegraphics[width=1\linewidth]
{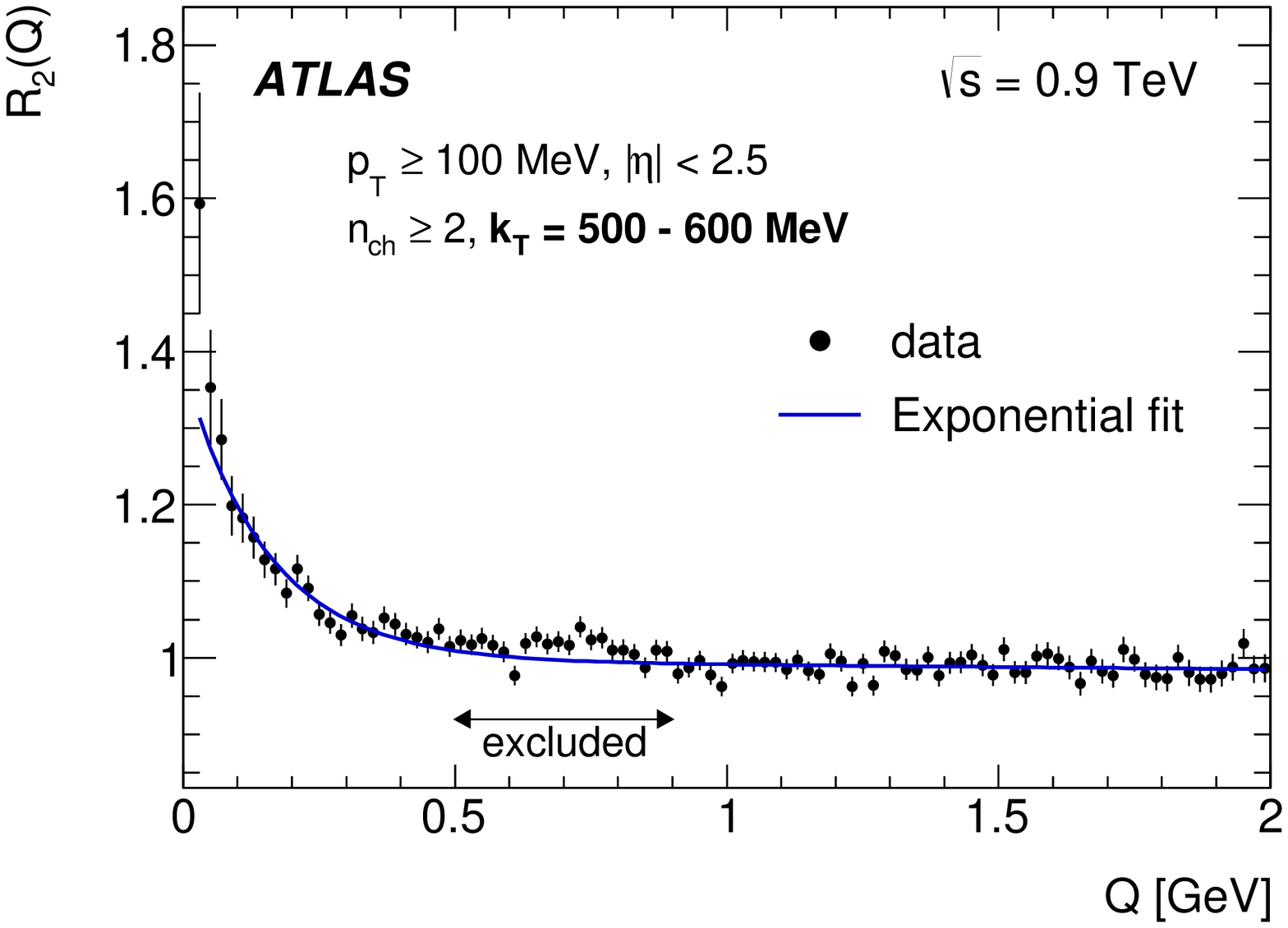}} 
(a) \\
\end{minipage}
\hfill
\begin{minipage}[h]{0.50\textwidth}
\center{\includegraphics[width=1\linewidth,height=0.75\textwidth]
{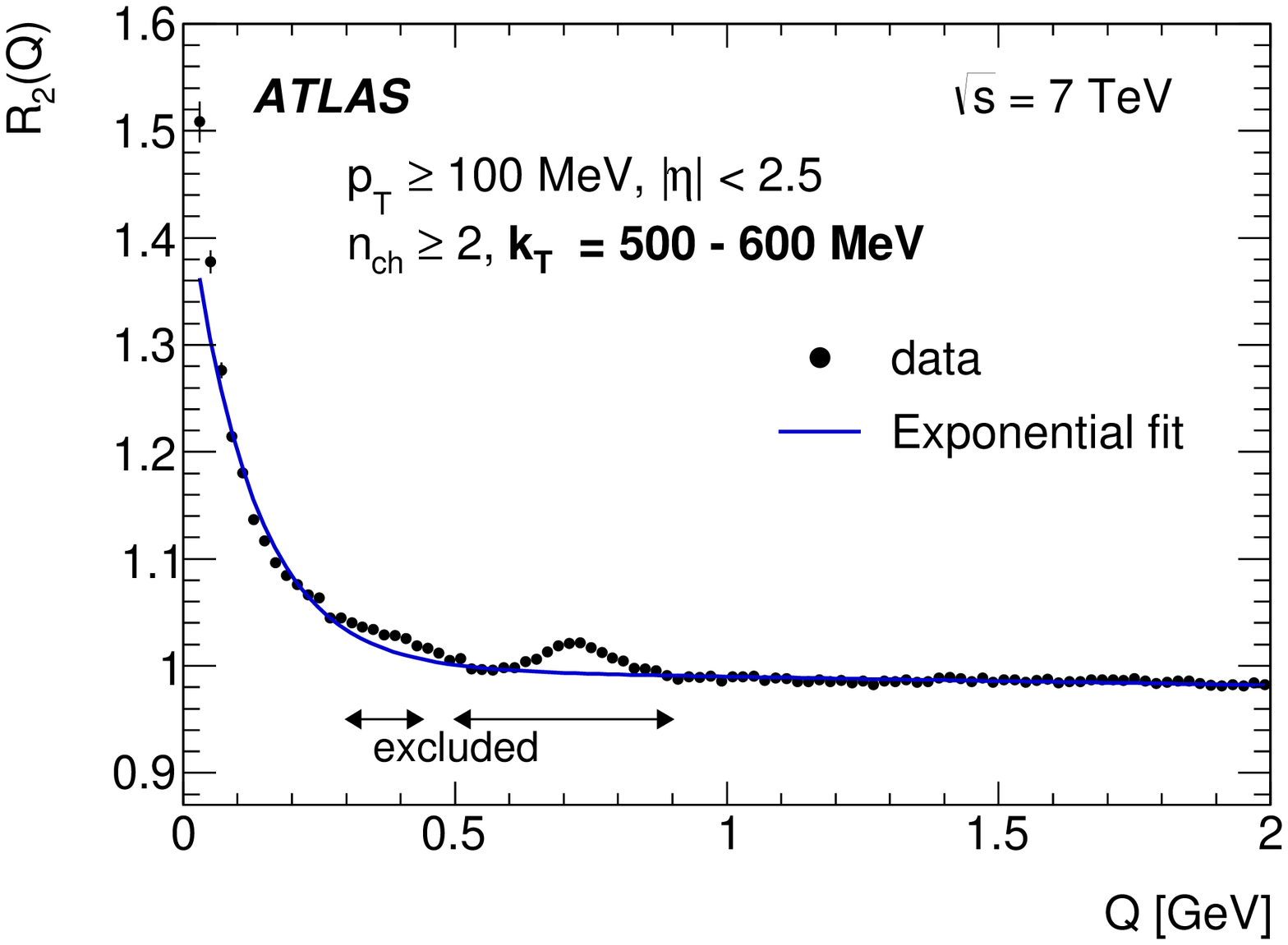}} 
(b)\\
\end{minipage}
 \vfill
 \begin{center}
 \begin{minipage}[h]{0.50\textwidth}
 \center{\includegraphics[width=1\linewidth]
{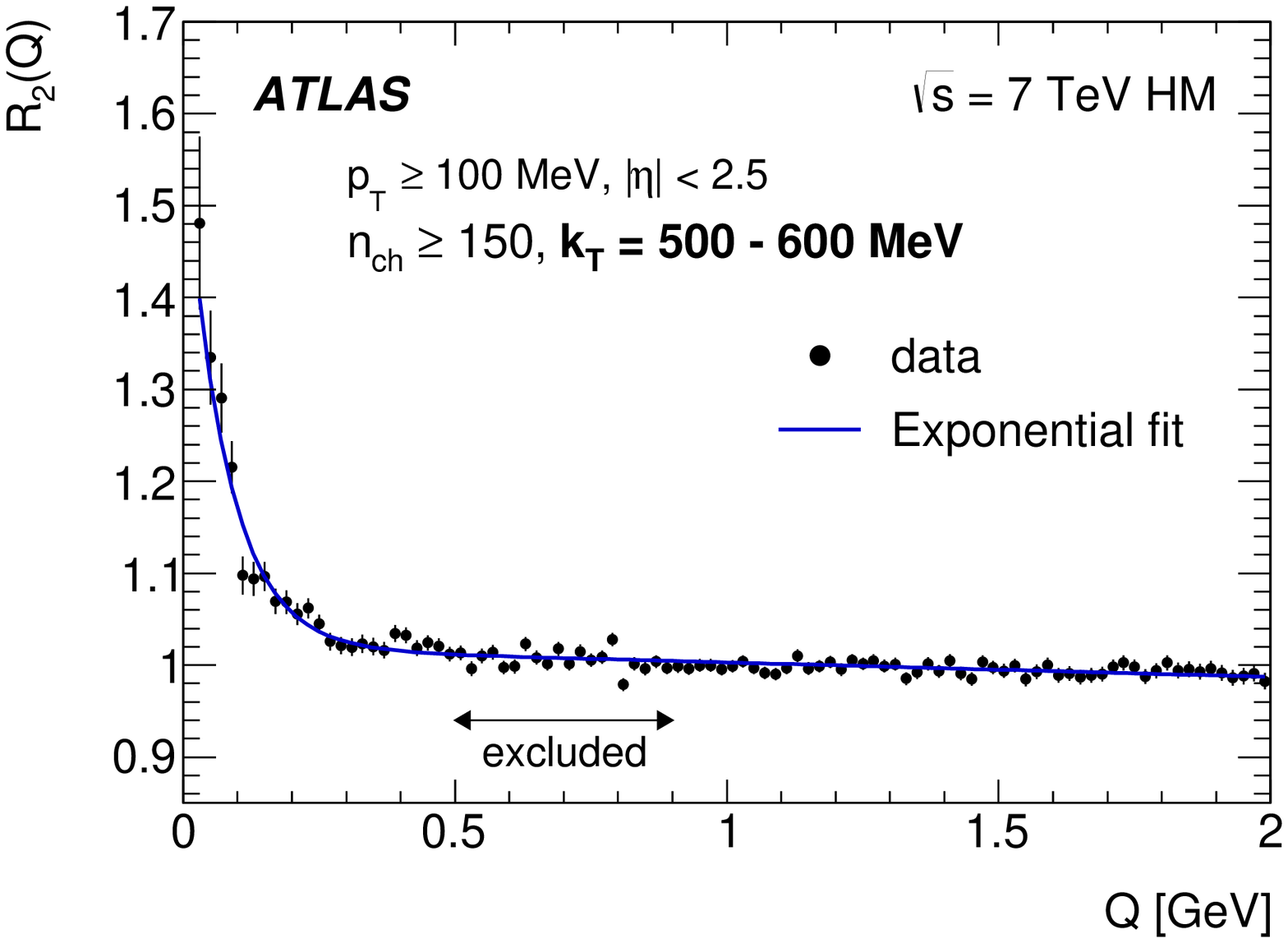}}  
(c) \\
\end{minipage}
\end{center}
\caption{
The two-particle double-ratio correlation function $R_2 (Q)$ for  charged particles in $pp$ collisions  for $500\le \kT < 600$ \MeV\ interval  at 
(a) $\sqrt{s}=$0.9~\TeV, 
(b) 7~\TeV\   
and 
 (c)  7~\TeV\   high-multiplicity events. 
 The average transverse momentum  $\kT$ of the particle pairs is defined 
as $\kT = |{\bf p}_{{\rm T}, 1} + {\bf p}_{{\rm T},  2}|/2$.  
 The lines show the  exponential fits.  
 The region  excluded from the fits is indicated. 
The  error bars represent the statistical uncertainties.
}
\label{Fig_4_1b}
\end{figure*}

As an example, the $R_2(Q)$ distributions for the  
$500\leq\kT\leq600$~\MeV\ interval for the 0.9~\TeV, 
7~\TeV\ and high-multiplicity 7~\TeV\ samples are shown in Fig.\ \ref{Fig_4_1b}
together with the results of the corresponding exponential fit. 
For the $R_2 (Q)$ correlation function measured at 7~\TeV\  
(see Fig.\ \ref{Fig_4_1b}(b)),   there is an indication that
 the Monte Carlo  simulation overestimates the production and decay of the $\omega$-meson 
 in the $Q$ region of $0.3$--$0.44$~\GeV.
 This region is thus excluded from the fit range 
for $\kT>500$~MeV bin 
results.

In the region most important for the BEC parameters,  the quality of the exponential fit is 
found to deteriorate as $\kT$ increases. 
This is due to the fact that at  large $\kT$ values, the characteristic BEC peak becomes steeper 
than the exponential function can accommodate.  
Despite the deteriorating fit quality, the behaviour of the fitted parameters is presented for 
comparison with previous experiments.

\begin{figure*}[t!]
\begin{minipage}[h]{0.50\textwidth} 
\center{\includegraphics[width=1\linewidth]
{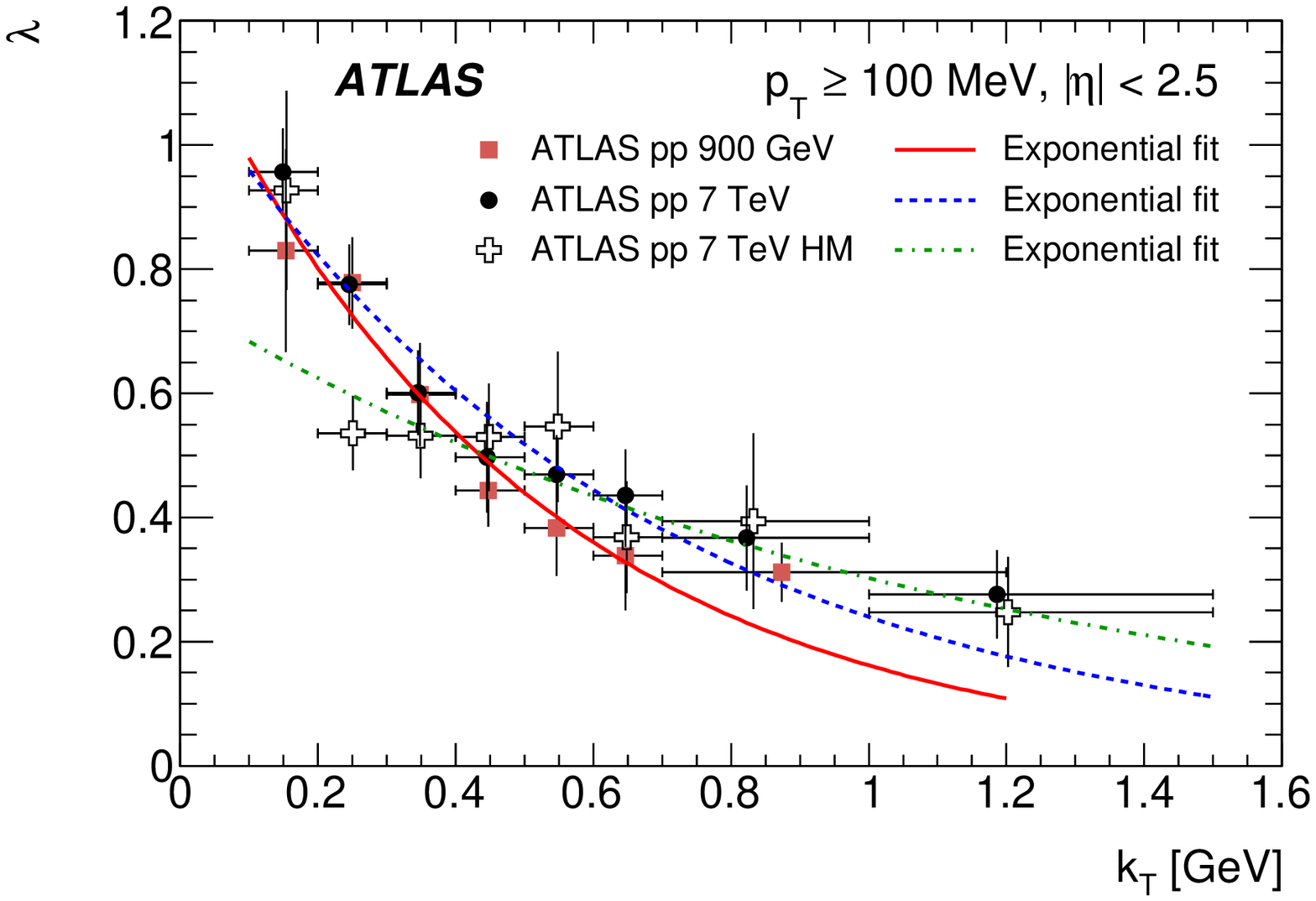}}
(a) \\
\end{minipage}
\hfill
\begin{minipage}[h]{0.50\textwidth}
\center{\includegraphics[width=1\linewidth]
{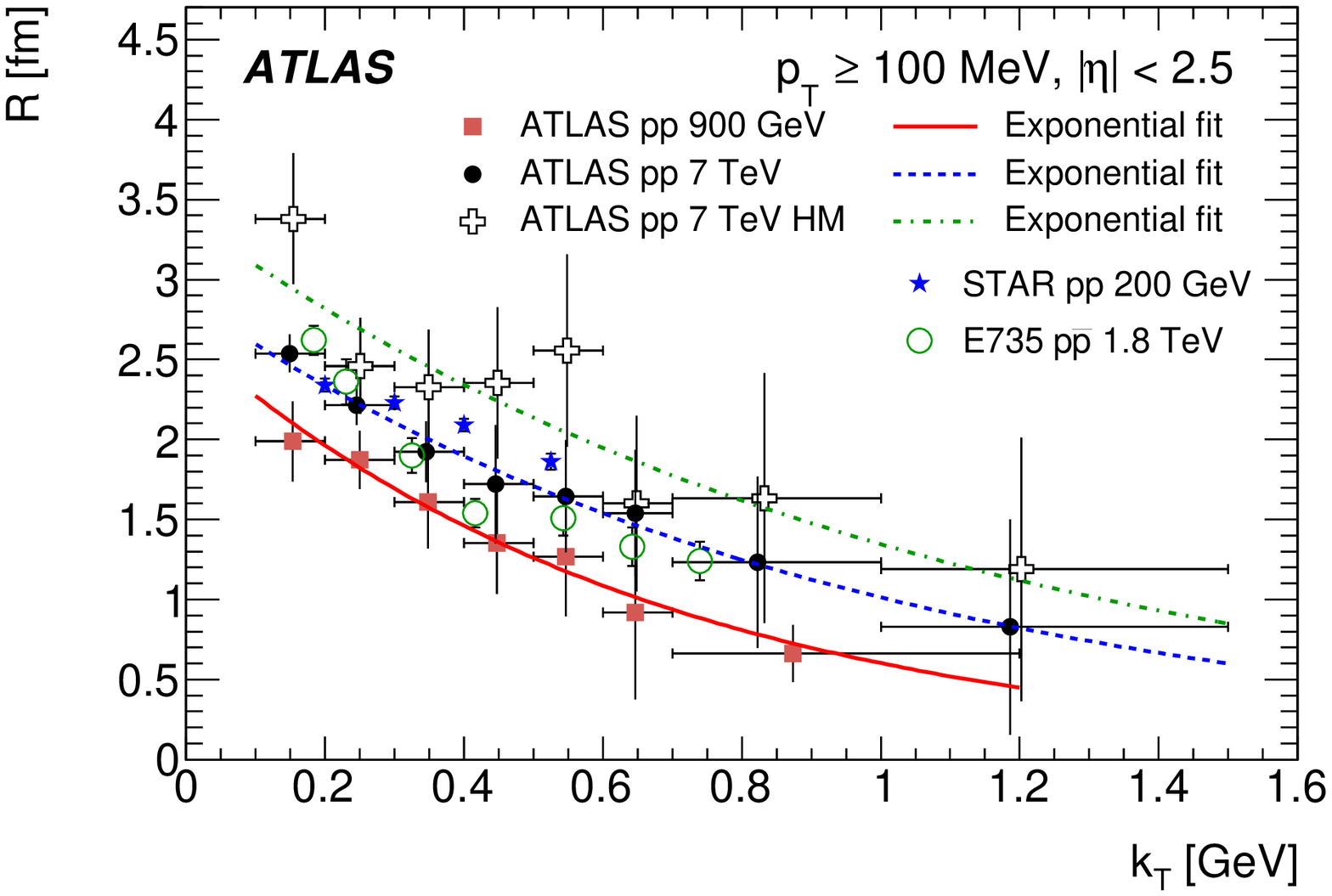}}
(b) 
\\
\end{minipage}
\caption{
The $\kT$ dependence of the fitted  parameters (a) $\lambda$  and  (b) $R$  obtained from the exponential fit
 to two-particle double-ratio 
 at $\sqrt{s}=$0.9~\TeV, 7~\TeV\  and 7~\TeV\  high-multiplicity events. 
The average transverse momentum  $\kT$ of the particle pairs is defined as $\kT = |{\bf p}_{{\rm T}, 1} + {\bf p}_{{\rm T},  2}|/2$.  
The  solid, dashed and 
dash-dotted curves are results of the exponential fits for 
0.9 ~\TeV,  7~\TeV\ and 7~\TeV\  high-multiplicity data, respectively.
 The results are compared to the corresponding measurements by 
the E735 experiment  
at the Tevatron \ct{e735-bec}, and  by 
the STAR experiment 
at RHIC \ct{star-bec200}. 
 The error bars  
represent the quadratic sum of the statistical and systematic uncertainties.
}
\label{Fig_4_e}
\end{figure*}

The fit values of the $\lambda$ and $R$ parameters are 
shown in Fig.\ \ref{Fig_4_e} as a function of $k_\mathrm{T}$.
The values of both $\lambda$ and $R$ decrease  with increasing $\kT$. 

The decrease of $\lambda$ with $\kT$ is well described by an  exponential  function, 
$\lambda (\kT) = \mu\, {\mathrm e}^{- \nu \kT}$.
 The  $\kT$ dependence of the $R$ parameter  is also found to follow  an exponential  decrease,  
$R (\kT) = \xi\, {\mathrm e}^{- \kappa \kT}$. 
The shapes of the $\kT$ dependence  are similar  for the 7~\TeV\  and  the 7~\TeV\  high-multiplicity data.
The results of the fits are presented in Table \ref{tab:parameters}. 

In  Fig.\  \ref{Fig_4_e}(b), 
the $\kT$ dependence of the $R$ parameter is compared to the measurements 
performed by the E735 \ct{e735-bec}  
and the STAR \ct{star-bec200} experiments with mixed-event reference 
samples.  
These earlier results were obtained from Gaussian fits to the single-ratio 
correlation functions and therefore the values of the measured  radius  
parameters
 are multiplied by  $\sqrt{\pi}$  as discussed in Sect.  \ref{sect:params}.
 The values of  the  parameters  are observed to be energy-independent 
within the uncertainties.

\begin{figure*}[t!]
\begin{minipage}[h]{0.50\textwidth} 
\center{\includegraphics[width=1\linewidth]
{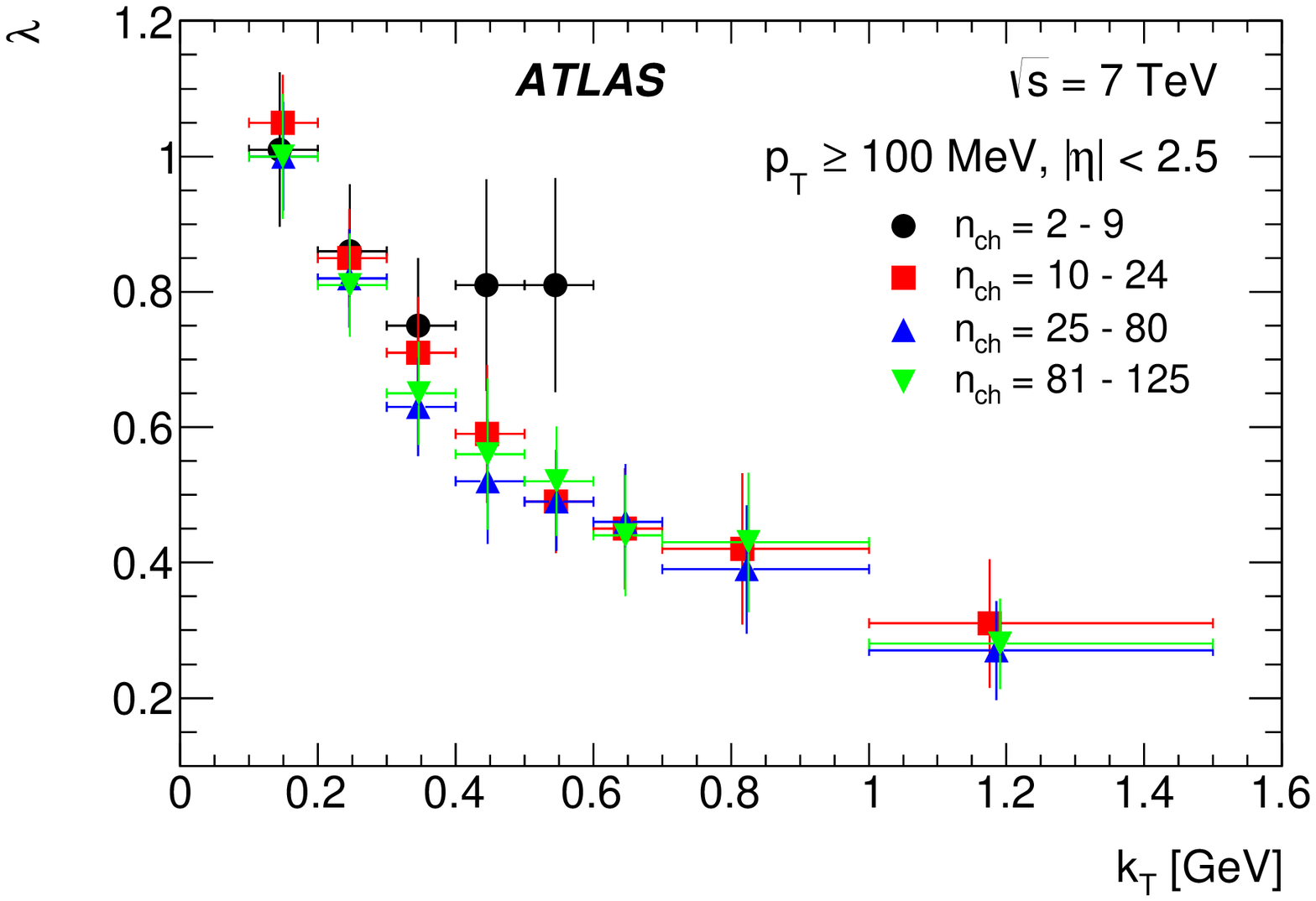}}
(a) \\
\end{minipage}
\hfill
\begin{minipage}[h]{0.50\textwidth}
\center{\includegraphics[width=1\linewidth]
{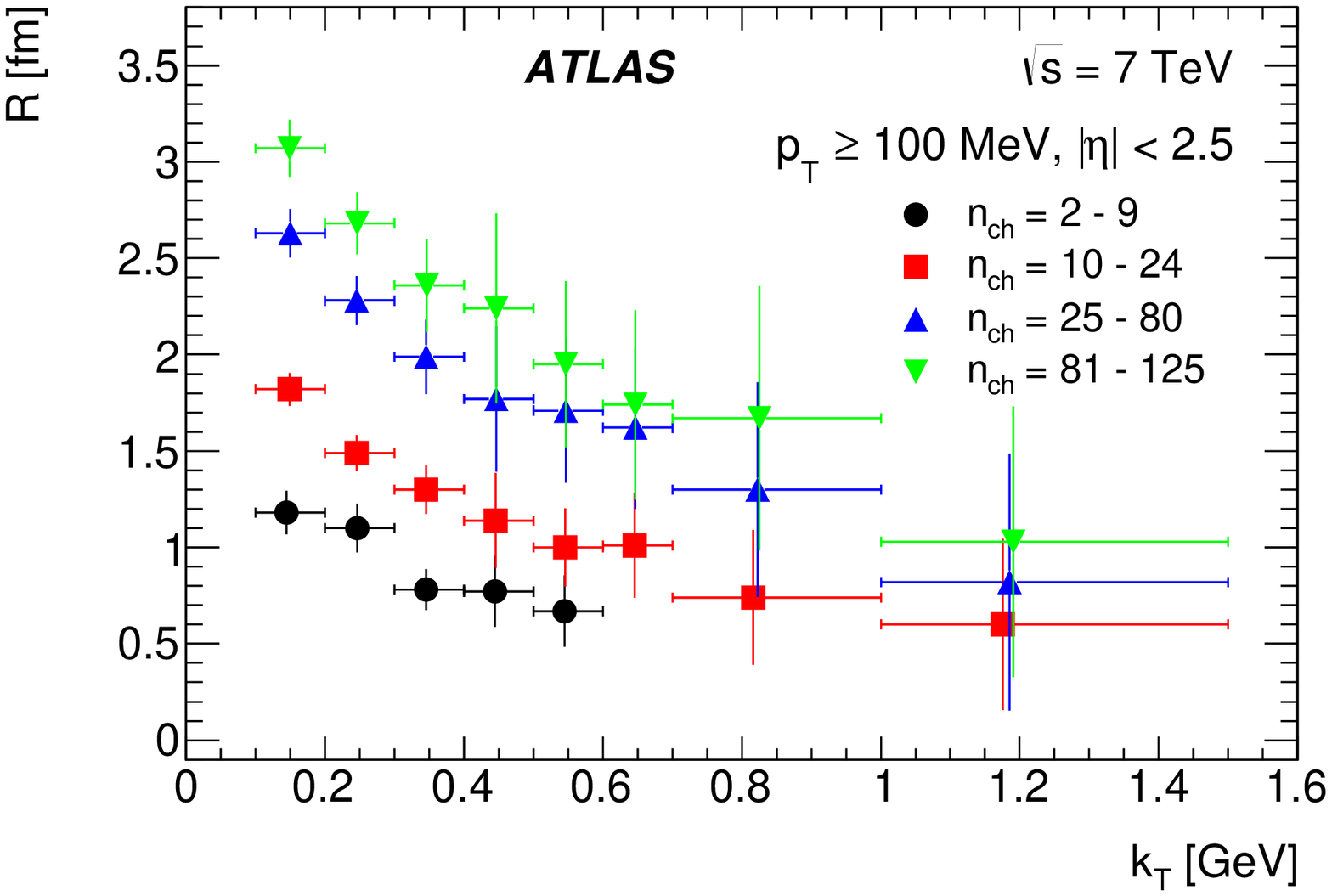}}
(b) \\
\end{minipage}
\caption{
 The $\kT$ dependence of the fitted
 parameters (a) $\lambda$  and (b) $R$  
obtained from the exponential fit to the two-particle double-ratio correlation function $R_2 (Q)$ 
 at $\sqrt{s}=$ 7~\TeV\   
 for the different 
multiplicity regions: 
$2\le n_{\rm ch} \le 9$ (circles),        
 $10 \le n_{\rm ch} \le 24$ (squares),  
$25 \le n_{\rm ch} \le 80$ (triangles) 	    
 and 
 $81 \le n_{\rm ch} \le 125$ (inverted
triangles). 
The average transverse momentum  $\kT$ of the particle pairs is defined as $\kT = |{\bf p}_{{\rm T}, 1} + {\bf p}_{{\rm T},  2}|/2$.  
 The error bars  represent the quadratic sum of the statistical and systematic uncertainties.
}
\label{Fig_kt_nch}
\end{figure*}

In  Fig.\ \ref{Fig_kt_nch},
the $\kT$ dependence of $\lambda$ and $R$, 
obtained for the 7~\TeV\ data, 
is also studied  in 
 various multiplicity regions:
$2 \le n_{\rm ch} \le 9$;  $10 \le n_{\rm ch} \le 24$;  $25 \le n_{\rm ch} 
\le 80$; and $81 \le n_{\rm ch} \le 125$.  
 The decrease of $\lambda$ with $\kT$ is nearly independent of 
multiplicity
for   $n_{\rm ch} > 9$ and the same as for the inclusive case. For $n_{\rm ch} \leq 9$ no conclusions can 
be drawn due to the 
large uncertainties.  
The   $R$-parameter decreases with  $\kT$ and exhibits an increase with 
increasing multiplicity as was observed for the fully inclusive case.

\clearpage

\clearpage

\section{Summary  and conclusions}

 The   two-particle  
Bose--Einstein correlations 
of like-sign  hadrons with $\pT >$~100~\MeV\ and $|\eta|<$~2.5
 produced  in $pp$ collisions  recorded by the ATLAS 
detector at 0.9 and 7~\TeV\  at the CERN Large Hadron Collider  are studied.
 In addition to minimum-bias data,  high-multiplicity 
data recorded at  7~\TeV\   using a dedicated    
 trigger are investigated.
The integrated luminosities are  about 7~$\mu$b$^{-1}$,  190~$\mu$b$^{-1}$  and   12.4~nb$^{-1}$ 
 for   0.9~\TeV,  7~\TeV\ minimum-bias  and 7~\TeV\  high-multiplicity  
 data samples,   respectively.

The studies were 
performed
using the double-ratio   correlation function.  
In the double-ratio method,   the single-ratio correlation function obtained 
from the data is divided   by a similar single-ratio calculated using Monte  Carlo events, 
which do not have BEC effects.
 The  reference sample  for each of the two  single-ratios is constructed from 
unlike-sign charged-particle pairs.
 
 A clear signal  
of  Bose--Einstein correlations is
observed in the region of small   four-momentum difference.
  To quantitatively characterize the BEC effect, Gaussian and
 exponential parametrizations are fit to the measured correlation
 functions.
 As observed in 
 studies performed by other experiments, 
 the Gaussian 
parameterization provides a poor description of the   BEC-enhanced region and 
 hence  the exponential parameterization  
is used  for the final results.  

 The BEC parameters   are  studied   as a function of the  charged-particle multiplicity and 
the transverse momentum of the particle pair.
 A decrease of the correlation strength $\lambda$ along with an increase 
of the correlation  source size 
parameter
$R$ are found with increasing  charged-particle 
multiplicity.
 On the other hand no dependence of $R$ on the 
centre-of-mass energy of $pp$ collisions is observed.
 For the first   time a saturation of the   source size 
 parameter 
is observed   for  multiplicity $n_{\rm ch}\geq 55$.
The correlation strength $\lambda$ and the  source size 
 parameter 
 $R$ are found to  decrease with increasing average transverse momentum of 
a pair.
The study of BEC in ($n_{\rm ch},
\kT$) bins at 7~\TeV\ shows a decrease of the $R$ parameter with $\kT$ for different 
multiplicity ranges, while the $R$ values increase with multiplicity. 
The $\lambda$ parameter is found to decrease with $\kT$ independently of 
the multiplicity range.
These  resemble the dependences for the inclusive case at 7~\TeV\ for 
minimum-bias and 
high-multiplicity 
data. 

 A comparison 
is made to the measurements by other experiments at the same and lower energies where possible. 
 The measurements presented here  
complement the earlier measurements
 by  extending  the studies to higher multiplicities and transverse momenta. 
 This   has allowed a first observation of a saturation in the magnitude of the   source radius   
parameter
 at high charged-particle multiplicities, and 
confirms the  exponential decrease, 
 observed in previous measurements   of  the 
 radius  
parameters
  with  increasing pair transverse  momenta.

\newpage
\bigskip
\bigskip
\noindent
{\large\bf Acknowledgments}
\bigskip
 
We are thankful to W.~Metzger for his input to this paper.
We thank CERN for the very successful operation of the LHC, as well as the
support staff from our institutions without whom ATLAS could not be
operated efficiently.

We acknowledge the support of ANPCyT, Argentina; YerPhI, Armenia; ARC,
Australia; BMWFW and FWF, Austria; ANAS, Azerbaijan; SSTC, Belarus; CNPq and\linebreak 
FAPESP,
Brazil; NSERC, NRC and CFI, Canada; CERN; CONICYT, 
Chile; CAS, MOST and NSFC, China; COLCIENCIAS, 
Colombia; MSMT CR, MPO CR and VSC CR, Czech Republic;
DNRF, DNSRC and Lundbeck Foundation, Denmark; 
EPLANET, ERC and NSRF, European Union;
IN2P3-CNRS, CEA-DSM/IRFU, France; GNSF, Georgia; BMBF, 
DFG, HGF, MPG and AvH Foundation, Germany; GSRT and 
NSRF, Greece; RGC, Hong Kong SAR, China; ISF, MINERVA, GIF, 
I-CORE and Benoziyo Center, Israel; INFN, Italy; MEXT and JSPS, 
Japan; CNRST, Morocco; FOM and NWO, Netherlands; 
BRF and RCN, Norway; MNiSW and NCN, Poland;\linebreak 
 GRICES and FCT, Portugal; MNE/IFA, Romania; 
MES of Russia and ROSATOM, Russian Federation; JINR;\linebreak  
MSTD, Serbia; MSSR, Slovakia; ARRS and MIZ\v{S}, Slovenia; DST/NRF, South Africa;
MINECO, Spain; SRC and Wallenberg Foundation, Sweden; SER, SNSF and Cantons of
Bern and Geneva, Switzerland; NSC, Taiwan; TAEK, Turkey; STFC, the Royal
Society and Leverhulme Trust, United Kingdom; DOE and NSF, United States of
America.

The crucial computing support from all WLCG partners is acknowledged
gratefully, in particular from CERN and the ATLAS Tier-1 facilities at
TRIUMF (Canada), NDGF (Denmark, Norway, Sweden), CC-IN2P3 (France),
KIT/GridKA (Germany), INFN-CNAF (Italy), NL-T1\linebreak  
(Netherlands), PIC (Spain),
ASGC (Taiwan), RAL (UK) and BNL (USA) and in the Tier-2 facilities
worldwide.

\onecolumn 
\clearpage 
\begin{flushleft}
{\Large The ATLAS Collaboration}

\bigskip

G.~Aad$^{\rm 84}$,
B.~Abbott$^{\rm 112}$,
J.~Abdallah$^{\rm 152}$,
S.~Abdel~Khalek$^{\rm 116}$,
O.~Abdinov$^{\rm 11}$,
R.~Aben$^{\rm 106}$,
B.~Abi$^{\rm 113}$,
M.~Abolins$^{\rm 89}$,
O.S.~AbouZeid$^{\rm 159}$,
H.~Abramowicz$^{\rm 154}$
% ...
\bigskip
\\
$^{1}$ Department of Physics, University of Adelaide, Adelaide, Australia\\
% ...
$^{11}$ Institute of Physics, Azerbaijan Academy of Sciences, Baku, Azerbaijan\\
% ...
$^{84}$ CPPM, Aix-Marseille Universit{\'e} and CNRS/IN2P3, Marseille, France\\
% ...
$^{89}$ Department of Physics and Astronomy, Michigan State University, East Lansing MI, United States of America\\
% ...
$^{106}$ Nikhef National Institute for Subatomic Physics and University of Amsterdam, Amsterdam, Netherlands\\
% ...
$^{112}$ Homer L. Dodge Department of Physics and Astronomy, University of Oklahoma, Norman OK, United States of America\\
$^{113}$ Department of Physics, Oklahoma State University, Stillwater OK, United States of America\\
% ...
$^{116}$ LAL, Universit{\'e} Paris-Sud and CNRS/IN2P3, Orsay, France\\
% ...
$^{152}$ Institute of Physics, Academia Sinica, Taipei, Taiwan\\
% ...
$^{154}$ Raymond and Beverly Sackler School of Physics and Astronomy, Tel Aviv University, Tel Aviv, Israel\\
% ...
$^{159}$ Department of Physics, University of Toronto, Toronto ON, Canada\\
% ...
\end{flushleft}


\end{document}